\documentclass[twocolumn]{aastex63}

\newcommand{\tng}{IllustrisTNG}
\newcommand{\arepo}{\textsc{Arepo}}

\received{}
\revised{}
\accepted{}
\submitjournal{ApJ}

\shorttitle{Dispersion profiles in TNG}
\shortauthors{Bose \& Loeb}
\graphicspath{{./}{figures/}}

\begin{document}

\title{Measuring the mass and concentration of dark matter halos\\ from the velocity dispersion profile of their stars}

\correspondingauthor{Sownak Bose}
\email{sownak.bose@cfa.harvard.edu}

\author[0000-0002-0974-5266]{Sownak Bose}
\affil{Center for Astrophysics $\vert$ Harvard \& Smithsonian \\
60 Garden St. \\
Cambridge, MA 02138, USA}

\author{Abraham Loeb}
\affil{Center for Astrophysics $\vert$ Harvard \& Smithsonian \\
60 Garden St. \\
Cambridge, MA 02138, USA}



\begin{abstract}

We use the IllustrisTNG (TNG) simulations of galaxy formation to measure the velocity dispersion profiles of dark matter and stars in Milky Way-mass, galaxy group, and cluster-scale dark matter halos. The mean profiles calculated from both tracers are similar in shape, exhibiting a large halo-to-halo scatter around the average profile. The so-called ``splashback'' radius demarcates the outer boundary of the halo, and manifests as a kink in the velocity dispersion profile, located on average between $\sim 1.0-1.5r_{200m}$, where $r_{200m}$ is the radius within which the density of the halo equals 200 times the background density of the universe. We find that this location may also be identified as the radius at which the (stacked)  dispersion profile drops to 60\% of its peak value (for line-of-sight motions in TNG halos). We further show that the scatter in the dispersion profiles may be attributed to the variations in the assembly history of the host halos. In particular, this segregates the profile into two regimes: one within $\sim0.1r_{200m}$, where the scatter is set by the early assembly history of the halo, and the other beyond this radius where the scatter is influenced more strongly by its late-time assembly. Finally, we show that a two-parameter model can be used to fit the measured velocity dispersion profiles and the fit parameters can be related directly to two fundamental halo properties: mass and concentration. We describe a simple model which allows us to express the stellar velocity dispersion profile in terms of these halo properties only.

\end{abstract}



\section{Introduction} 
\label{sec:intro}

Dark matter halos are the fundamental units of cosmic structure formation. These objects are formed through the assembly of dark matter particles, in which initially overdense regions of the universe collapse through gravitational instability. In hierarchical structure formation, low-mass dark halos form first, while more massive structures form gradually through mergers, combined with the smooth accretion of dark matter. Eventually, baryonic matter settles into the gravitational potential wells of these halos, leading to the eventual cooling of gas into a star-forming state, and the subsequent production of stars and black holes. 

Given the chronology of structure formation, understanding the assembly and present-day mass distribution of dark matter halos is an important first step in developing a comprehensive theory of galaxy formation. $N$-body simulations of collisionless `cold' dark matter agree that the mass distribution within dark halos takes on a universal shape, most conveniently paramterized by the Navarro-Frenk-White (NFW) density profile \citep{Navarro1996,Navarro1997,Wang2020}. The NFW density profile is defined such that for halos of fixed mass, the density distribution depends on only one additional parameter, the so-called `concentration' of the halo -- a measure of how centrally peaked the density profile is. Concentration may also be interpreted  as a measure of the formation epoch of the halo, with low-mass, older halos exhibiting larger concentrations than high-mass objects that have assembled more recently in cosmic history \citep[e.g.][]{Bullock2001,Wechsler2002,Zhao2003,Ludlow2013}.

The concept of halo mass is not without ambiguity. Dark matter halos exhibit irregular shapes, and quantifying the mass of a halo then becomes an exercise in determining a suitable boundary demarcating its dark matter content. The most commonly-adopted definitions of halo boundaries originate from the spherical collapse model \citep{Gunn1972}, in which halos condense out of initial overdensities that decouple from the background expansion through their own gravity; the resulting extent of the collapsed object is then determined using the virial theorem. This condition then defines halos as virialized objects with a mean enclosed density equal to $178$ times the critical density of the universe. This condition applies specifically to an Einstein de-Sitter universe (with $\Omega_m=1$); the extension to arbitray cosmologies is more involved and requires numerical solutions \citep[e.g.][]{Eke1996,Bryan1998,Rubin2013}. Nevertheless, these virialization conditions act as convenient criteria that can be used to identify (spherical) halos from groups of particles in cosmological simulations of structure formation, where an overdensity closer to 200 times the critical (or, sometimes, the mean) density is assumed. The corresponding ``virial'' radius is  denoted as $r_{200c}$ ($r_{200m}$).

Visual depictions of halos in simulations show them to be anything but spherical. In particular, the equilibrium state of halos is, in some ways, a function of radial distance from the halo center. The inner cores of halos assemble earlier and are typically more relaxed. On the other hand, the exterior portions of halos (particularly those of galaxies more massive than the Milky Way) may still be accreting new material in the form of diffuse dark matter and merging halos, and do not show an obvious separation from the cosmological background. This has led to several authors advocating for the ``splashback radius'' as a more natural definition of a halo boundary \citep[e.g.][]{Fillmore1984,Bertschinger1985,Diemer2014,Adhikari2014,More2015}, identified as a caustic in the density profile in the outskirts of halos. Physically speaking, the splashback definition may be more robustly defined as the (smoothed) average of the apocentric radii of all particles in the halo \citep{Diemer2017}. Defined as such, the splashback radius then separates material that is infalling from that which is in orbit within the gravitational potential of the halo.

A number of recent programs have targeted the identification of the splashback feature in observations. The majority of these efforts have been focused in the regime of rich galaxy clusters, in which the comparatively large number of tracers enables a more faithful detection of the caustic in the (projected) number density profile and cross-correlation functions of member galaxies \citep[e.g.][]{More2016,Patej2016,Baxter2017,Chang2018,Nishizawa2018,Shin2019,Zurcher2019,Murata2020,Tomooka2020}. Though the comparison between the observationally-inferred and theoretically-predicted splashback radii can be hindered by systematics \citep[e.g.][]{Busch2017}, there is now a considerable body of evidence that the outermost realms of massive galaxies are imprinted with a caustic splashback feature.

A particularly interesting feature of the splashback radius is its dependence on redshift, the accretion rate and environment of the halo \citep[e.g.][]{Diemer2017b,Mansfield2017} and perhaps even the underlying theory of gravity \citep{Adhikari2018}. The dependence on accretion rate is especially informative, as it influences the relationship between the splashback radius and the more conventionally defined virial radius derived from the spherical collapse model. Using cosmological $N$-body simulations, \cite{More2015} observe that in slowly accreting halos, the splashback feature occurs at around $\approx 1.2-1.5\, r_{200m}$, while in more rapidly accreting systems, in which the added mass causes particle orbits to ``turn around'' at smaller radii, this feature occurs closer to $\approx 0.8-1.0\, r_{200m}$. The dynamics of particle orbits, which after all defines the splashback radius, therefore retains memory of the accretion events that punctuate the history of the halo. 

Recent theoretical works have demonstrated that caustic features that demarcate natural halo boundaries may be measured in other quantities besides the radial density profile. For example, \cite{Fong2020} construct a new definition, the ``depletion radius'', by identifying the radial location of the minimum in the bias profile of halos (i.e. a measure of the overdensity profile of a given object). The location of this depletion radius is somewhat larger than the typical splashback radius (on average by a factor of two or greater), and coincides with the location of the maximum infall velocity around the halo.  Using high-resolution, zoom hydrodynamical simulations of Milky Way-mass analogs, \cite{Deason2020} showed that a series of caustics may also be identified in the radial velocity dispersion profile, with the innermost caustic (located around $\sim 0.6 r_{200m}$) defined by material that has undergone at least two pericentric passages, and which provides a means to define the ``edge'' of the Milky Way halo.

Unsurprisingly, the (radial) stellar velocity dispersion profile has been measured most comprehensively in the case of the Milky Way \citep[e.g.][]{Brown2010,Deason2012,Cohen2017}. The extension to  galaxies beyond the Milky Way has been more limited in terms of the maximum projected distance up to which the velocity dispersion profile has been measured, where the component being measured is the stellar motion along the line-of-sight \citep[e.g.][]{Tempel2006,Veale2018,Mogotsi2019}. Identifying caustics in the velocity dispersion profile in the most exterior portions of external galaxies may be enabled by stacking profiles of multiple objects, and by obtaining deeper spectra from future observational facilities.

The aim of this paper is to investigate the extent to which visible tracers in the halo--in particular, their stellar content--can be used to determine aspects relating to the mass and the formation history of the host dark matter halo. We make use of the IllustrisTNG cosmological, hydrodynamical simulations \citep{Pillepich2018,Nelson2018b} to measure the velocity dispersion profile of stars of objects ranging from the scale of Milky Way-mass halos to those of rich clusters of galaxies. We find that sharp breaks in the velocity dispersion profile, occurring at radii consistent with the expected splashback radius of these halos, are prominent in both the dark matter and stellar velocity dispersion profiles. Furthermore, we establish a connection between the shape of these profiles and the mass and concentration of the host halos, which opens up the possibility to infer these quantities from the velocity dispersion profile of galaxies measured in future observations. 

This paper is organized as follows. In Section~\ref{sec:numerical}, we describe the simulation set used in this work. Our main results are presented in Section~\ref{sec:results}, in which we establish the connection between the stellar velocity dispersion profiles and the assembly history of halos. We further demonstrate how the form of the stellar velocity dispersion can be used to determine the virial radius of the host dark matter halo. Finally, Section~\ref{sec:conclusions} provides a summary of our investigation.

\section{Numerical Methods}
\label{sec:numerical}
 
 First, we provide a brief description of the \tng{} simulation suite, which provides the computational domain analyzed in this paper.

\subsection{Simulations}
\label{sect:sims}

The \tng{} project is a suite of cosmological, magneto-hydrodynamical simulations of galaxy formation \citep{Pillepich2018b,Nelson2018a,Marinacci2018,Naiman2018,Springel2018}, carried out in periodic volumes with comoving lengths of 50, 100 and 300 Mpc (TNG50, TNG100, and TNG300, respectively, with only the two latter boxes used in this paper). Each simulation follows the co-evolution of dark matter and baryons, and has been run using the \arepo{} code \citep{Springel2010,Weinberger2019}, in which the equations of magneto-hydrodynamics governing gas elements are solved using an unstructured, Voronoi mesh. The Voronoi tessellation is adaptive in nature: regions with high gas density are resolved with many small cells as compared to more diffuse regions. This vastly increases the dynamical range achievable in any given simulation. 

The TNG model is the successor to the original Illustris galaxy formation model presented in \cite{Vogelsberger2013,Vogelsberger2014a}; a comprehensive list of all the changes introduced in the new version is provided in \cite{Pillepich2018}. TNG  incorporates a range of physical processes thought to be important to regulating the evolution of galaxies, including gas cooling and star formation; the seeding, growth and feedback resulting from supermassive black holes; the launching of galactic winds, as well as the influence of galactic-scale magnetic fields. A sequence of works have shown that this model is able to successfully reproduce a variety of properties of the real galaxy population as a function of cosmic time \citep{Pillepich2018b,Nelson2018a,Marinacci2018,Naiman2018,Springel2018}.

All simulation data used in this work (particle snapshots, halo and galaxy catalogs, and merger trees) have been made publicly available\footnote{\href{http://www.tng-project.org/data/}{http://www.tng-project.org/data/}} \citep{Nelson2018b}. The higher resolution TNG100 simulation consists of a periodic box of length $L_{{\rm box}} = 75\,h^{-1}$Mpc $\approx$ 100 Mpc, with 2$\times$1820$^3$ resolution elements corresponding to dark matter particles and gas cells. This corresponds to a mass resolution of $9.44\times10^5\,h^{-1}\,{\rm M}_\odot$ in baryons and $5.06\times10^6\,h^{-1}\,{\rm M}_\odot$ in dark matter. The maximum physical softening length of dark matter and star particles is set to $0.5\,h^{-1}$kpc. TNG300 simulates an even larger cosmological box of size $L_{{\rm box}} = 205\,h^{-1}$Mpc $\approx$ 300 Mpc with 2$\times$2500$^3$ resolution elements. The increased volume comes at the expense of more modest particle resolution: in particular, the mass resolution is $7.44\times10^6\,h^{-1}\,{\rm M}_\odot$ in baryonic matter and $3.98\times10^7\,h^{-1}\,{\rm M}_\odot$ in dark matter. The maximum physical softening length of dark matter and star particles is set to $1.0\,h^{-1}$kpc. TNG300 serves as our primary dataset for sampling the
regime of low-mass groups and rich galaxy clusters ($\gtrsim 10^{13}\,{\rm M}_\odot$), which, in particular, is the mass range of interest in this paper.

Both sets of simulations have been evolved until $z=0$, starting from initial conditions generated at $z=127$. The initial particle set is constructed by assuming cosmological parameters estimated by {\it Planck} \citep{Planck2016}: $\Omega_0 = 0.3089$ (total matter density), $\Omega_{\rm b} = 0.0486$ (baryon density), $\Omega_\Lambda = 0.6911$ (dark energy density), $H_0 = 67.74$ kms$^{-1}$Mpc$^{-1}$ (Hubble parameter) and $\sigma_8 = 0.8159$ (linear rms density fluctuation in a sphere of radius 8 $h^{-1}$ Mpc at $z=0$). 

\begin{figure*}
\centering
\includegraphics[width=0.475\textwidth]{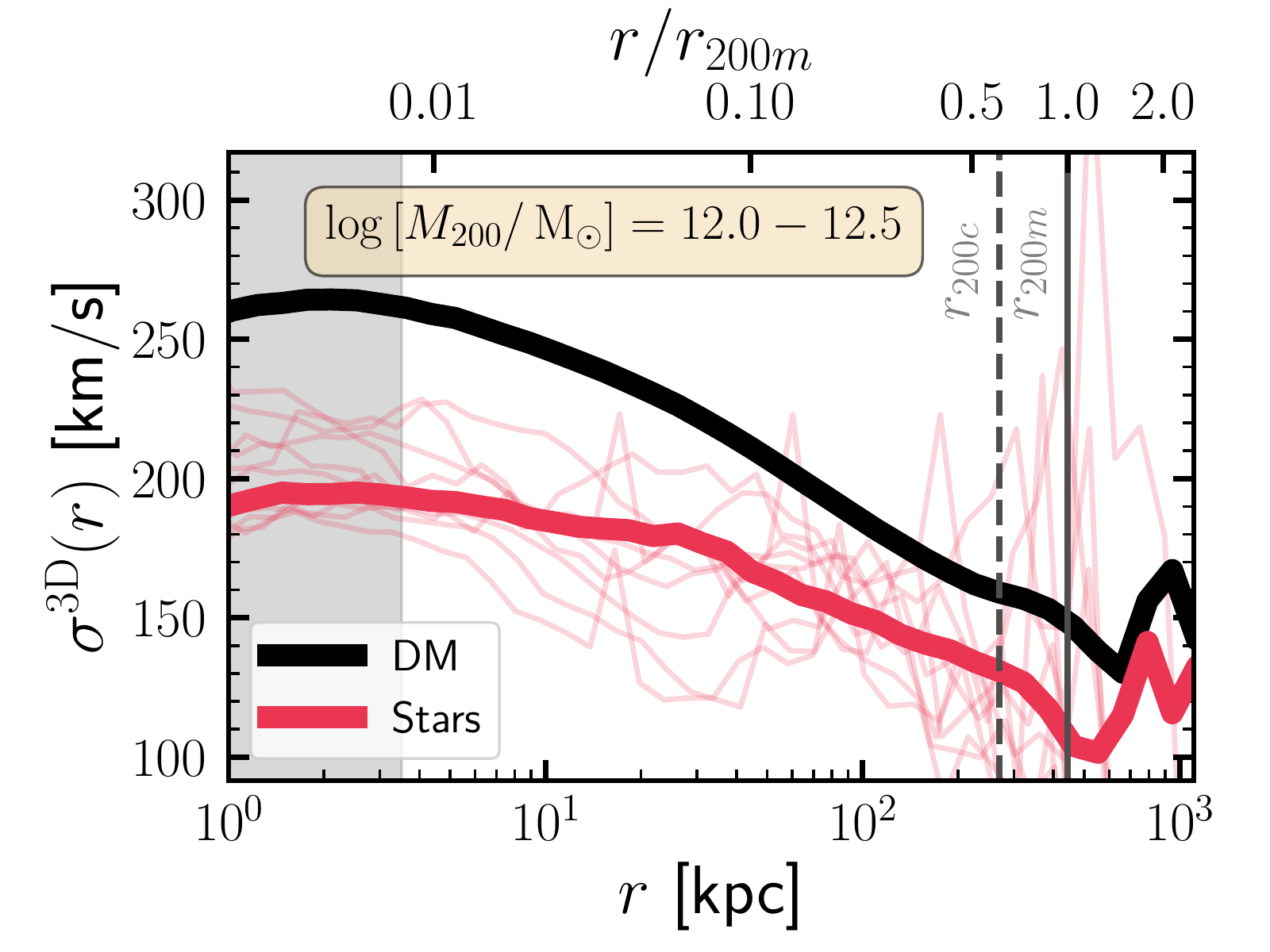}
\includegraphics[width=0.475\textwidth]{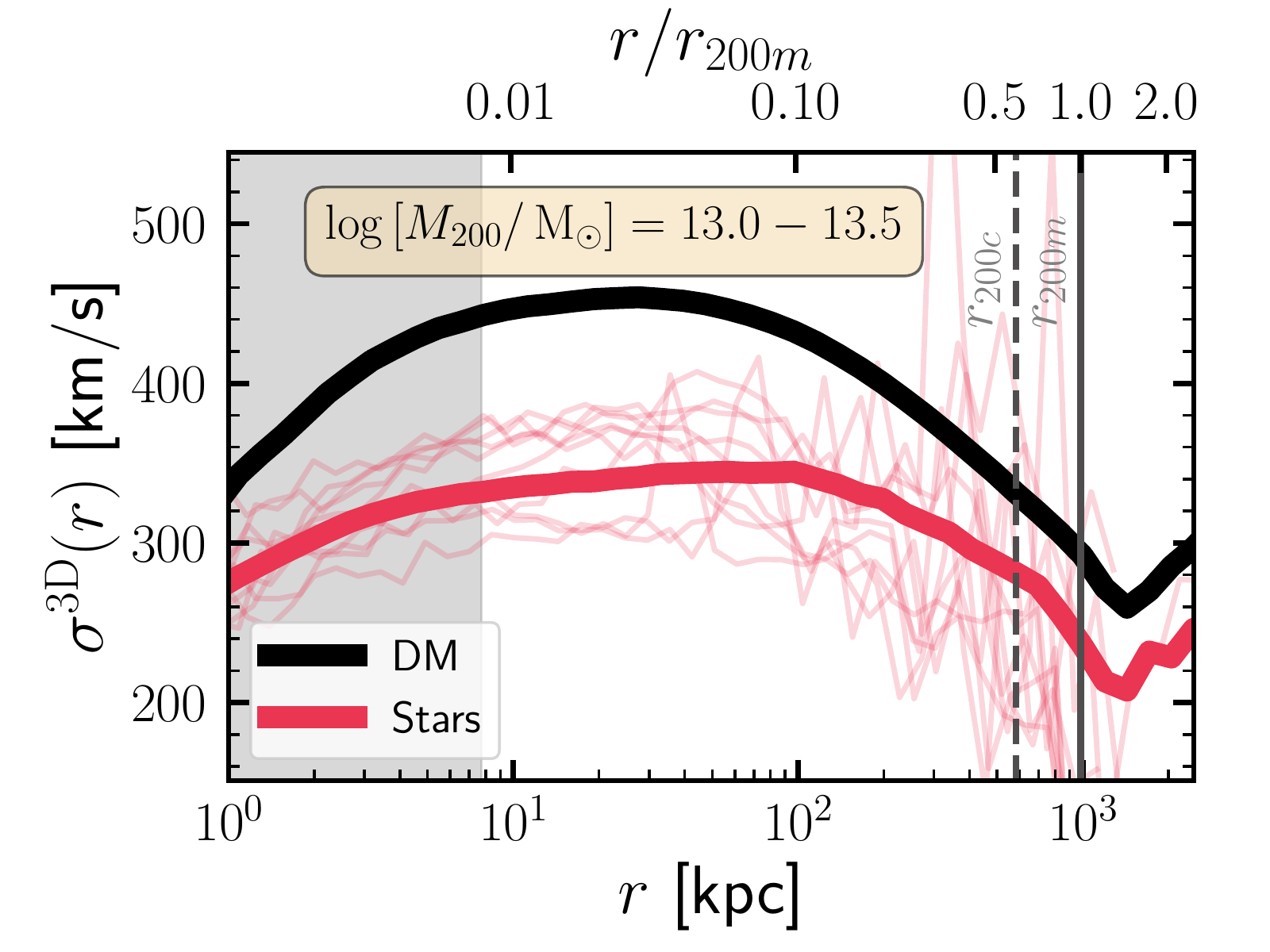}\\ 
\includegraphics[width=0.475\textwidth]{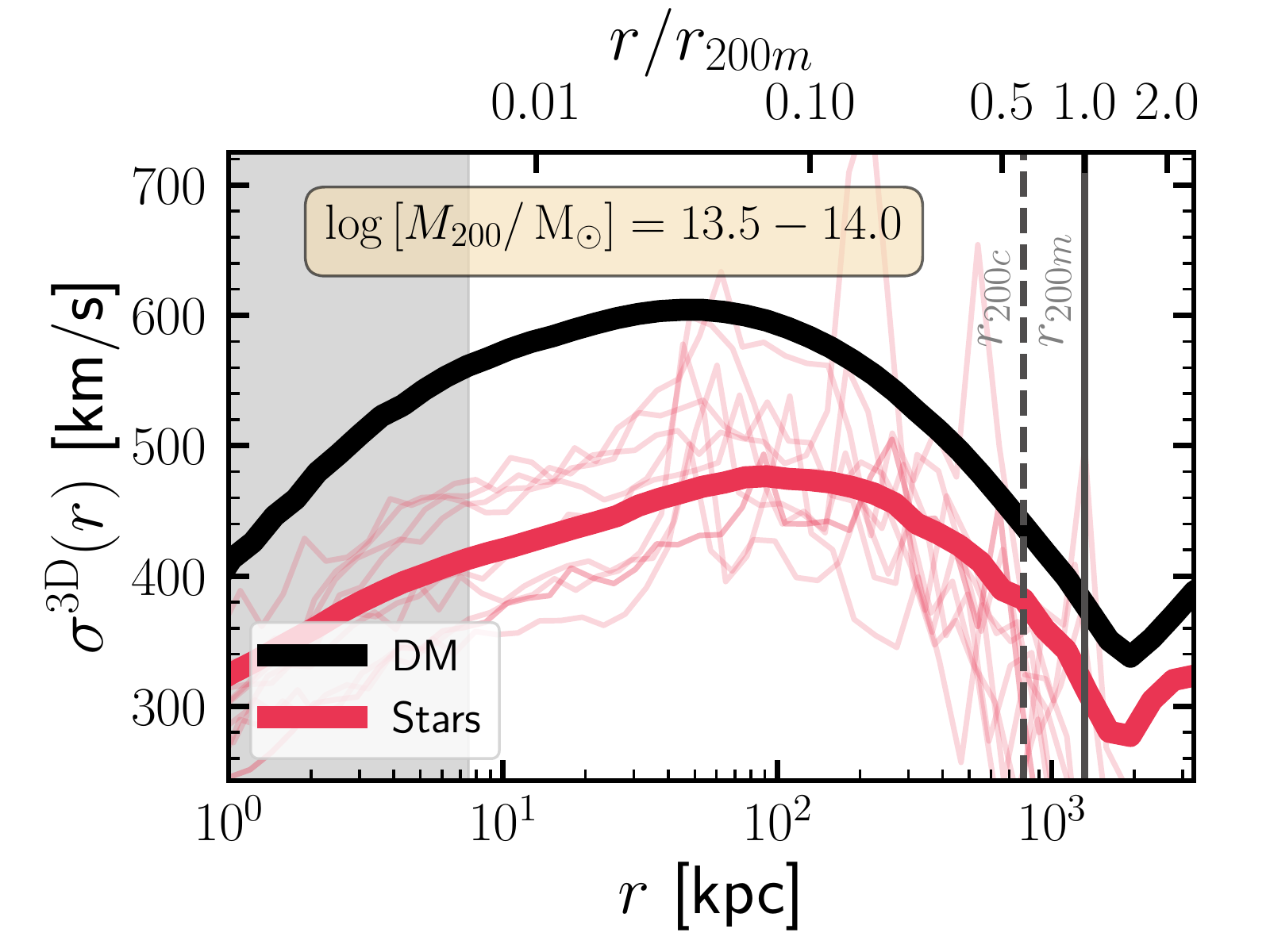}
\includegraphics[width=0.475\textwidth]{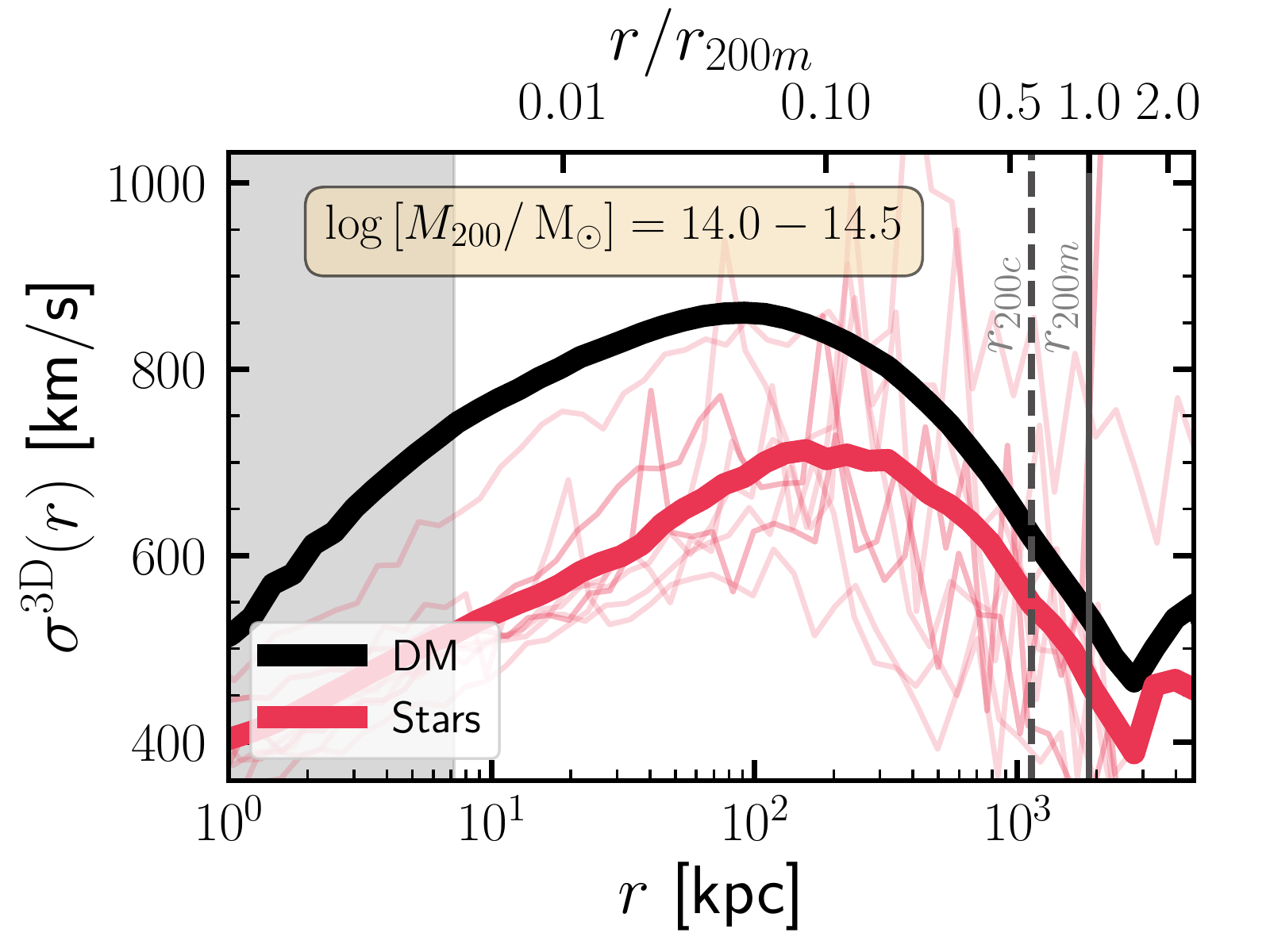}\\
\caption{The stacked 3D velocity dispersion profiles of dark matter (black) and star particles (red) in halos identified from TNG. The thick lines represent mean profiles, whereas the thin red curves show a selection of individual profiles from a subset of halos in each mass bin (as defined in the textbox within each sub-panel). The vertical dashed and solid lines, respectively, mark the averaged values of $r_{200c}$ and $r_{200m}$ for each mass bin. The shaded gray band is bounded by the convergence radius, below which the dispersion profiles can no longer be expected to have converged numerically. The shape of the velocity dispersion profile for stellar particles is similar to that of the dark matter, albeit with a lower amplitude.}
\label{fig:3d_disp}
\end{figure*}

\subsection{Halo identification}
\label{sec:identify}

Dark matter particles in all TNG simulations are first linked together using the `friends-of-friends' (FOF) algorithm \citep[e.g.][]{Davis1985}, providing an initial catalog of dark matter halos. This FOF algorithm connects dark matter particles separated by at most 0.2 times the mean interparticle separation to form groups; the \textsc{Subfind} algorithm \citep{Springel2001} is then used to identify gravitationally bound substructures within each group. In this work, we will focus on the properties of FOF groups as a whole, in particular, considering the set of dark matter and star particles associated with each FOF group. Throughout this paper, we will refer to the mass of a FOF group as $M_{200}$, which is the mass contained within the radius $r_{200c}$, the radius which encloses a mean density equal to 200 times the critical density of the universe at the redshift at which the halo is identified. We will also quote a second radius, $r_{200m}$, which is the radius within which the enclosed density of the halo is equal to 200 times the mean {\it background} density of the universe at that redshift. Defined in this way, $r_{200m}$ is always greater than $r_{200c}$ (for typical NFW-like halos, $r_{200m} \approx 1.6 r_{200c}$). Owing to its substantially larger volume, we use TNG300 as our primary data set for halos more massive than $\log\left[M_{200}/{\rm M}_\odot\right]\geq13.0$; for less massive halos ($\log\left[M_{200}/{\rm M}_\odot\right]=\left[12.0, 13.0\right]$), we resort to TNG100, which offers somewhat better mass and force resolution in this regime.

\section{Results}
\label{sec:results}

In the following subsections, we present the main results of our analysis. In Section~\ref{sec:v3d}, we showcase the diversity of dark matter and stellar velocity dispersion profiles (in 3D) measured in TNG halos. Section~\ref{sec:assembly} then establishes the connection between the diversity of these profiles and the assembly history of the halo. Finally, in Section~\ref{sec:model}, we present a simple model that describes the dependence of stellar velocity dispersion profile on physical properties of the host halo, such as its mass and concentration. 

\subsection{Dispersion profiles in three dimensions}
\label{sec:v3d}

We begin our investigation by considering the three dimensional velocity dispersion profiles of dark matter and star particles measured from TNG halos. Throughout this paper, we consider all particles within the spherical region enclosed by a radius $5r_{200m}$ from the center of the halo.

Figure~\ref{fig:3d_disp} shows the 3D velocity dispersion profiles in bins of halo mass (separate panels). The thick lines in each color shows the mean profile of the dark matter (black) and the stars (red). To provide an impression of the diversity in dispersion profiles, a subset of individual stellar velocity curves are represented by the thin red lines. Finally, the gray shaded region marks the ``convergence radius'': the regime below which the velocity profiles cannot be expected to have converged numerically given the numerical settings adopted in TNG. To determine this radius, we use the criterion described in \cite{Power2003}. Tests of the convergence of these profiles as a function of resolution are shown in Appendix~\ref{sec:convergence}.

The similarity in the shape of the dark matter and stellar velocity dispersion profiles is apparent; indeed, both curves show a prominent dip towards the exterior of the halo, typically occurring at around $\sim 1.2-1.5r_{200m}$. The location of this feature is commensurate with that of the so-called ``splashback'' radius of dark matter halos \citep[e.g.][]{Fillmore1984,Bertschinger1985,Adhikari2014,Diemer2014,More2015}, which demarcates the outermost caustic in the matter distribution. Indeed, a sharp rise in the velocity dispersion of material in the outskirts of halos is a hallmark of the region that separates infalling material from their surroundings.

While the stacked dispersion profiles for dark matter and stars differ in their amplitude, it is striking to note the extent to which the stellar distribution traces the overall shape of the underlying dark matter potential, particularly for halos in the mass range $\log \left[ M_{200}/{\rm M}_\odot\right] \leq 14.0$. On inspecting the dispersion profiles of individual halos, we notice several cases where the `kink' in the exterior of the profile is barely noticeable; this is typically the case for well-isolated halos, in which the velocity dispersion drops precipitously beyond $\sim r_{200m}$. Where there is a second, neighboring halo, the velocity dispersion profiles turn upwards as as the gravitational potential of the neighboring object starts to dominate. The stacked, averaged profiles show this effect clearly.

In subsequent sections, we divert our attention away from 3D velocity dispersion profiles and focus instead on line-of-sight velocity dispersions as may be measured with deep measurements of galaxy spectra.

\subsection{The effect of assembly history on velocity dispersion profiles}
\label{sec:assembly}

In this subsection, we establish the connection between the diversity of velocity dispersion profiles and the assembly history of halos in \tng{}.

\begin{figure*}
\centering
\includegraphics[width=0.9\textwidth]{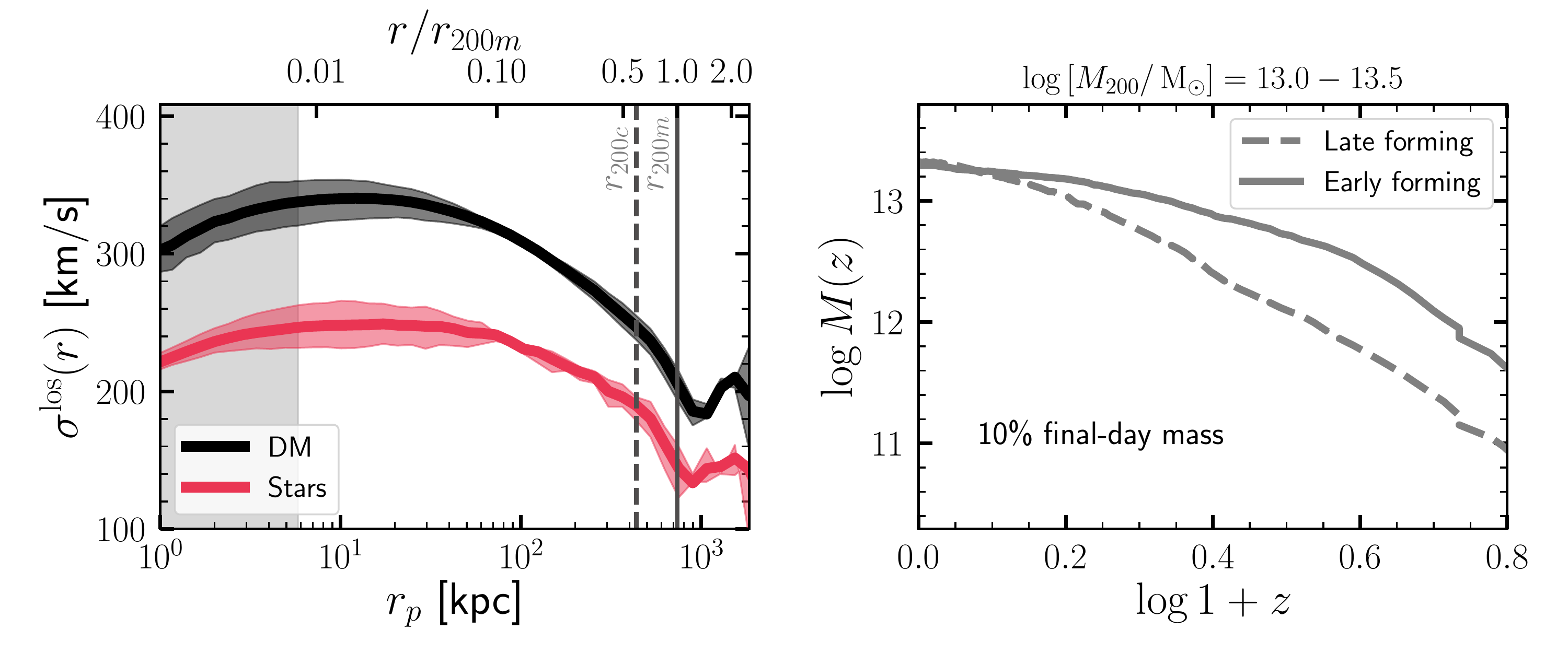} \\
\includegraphics[width=0.9\textwidth]{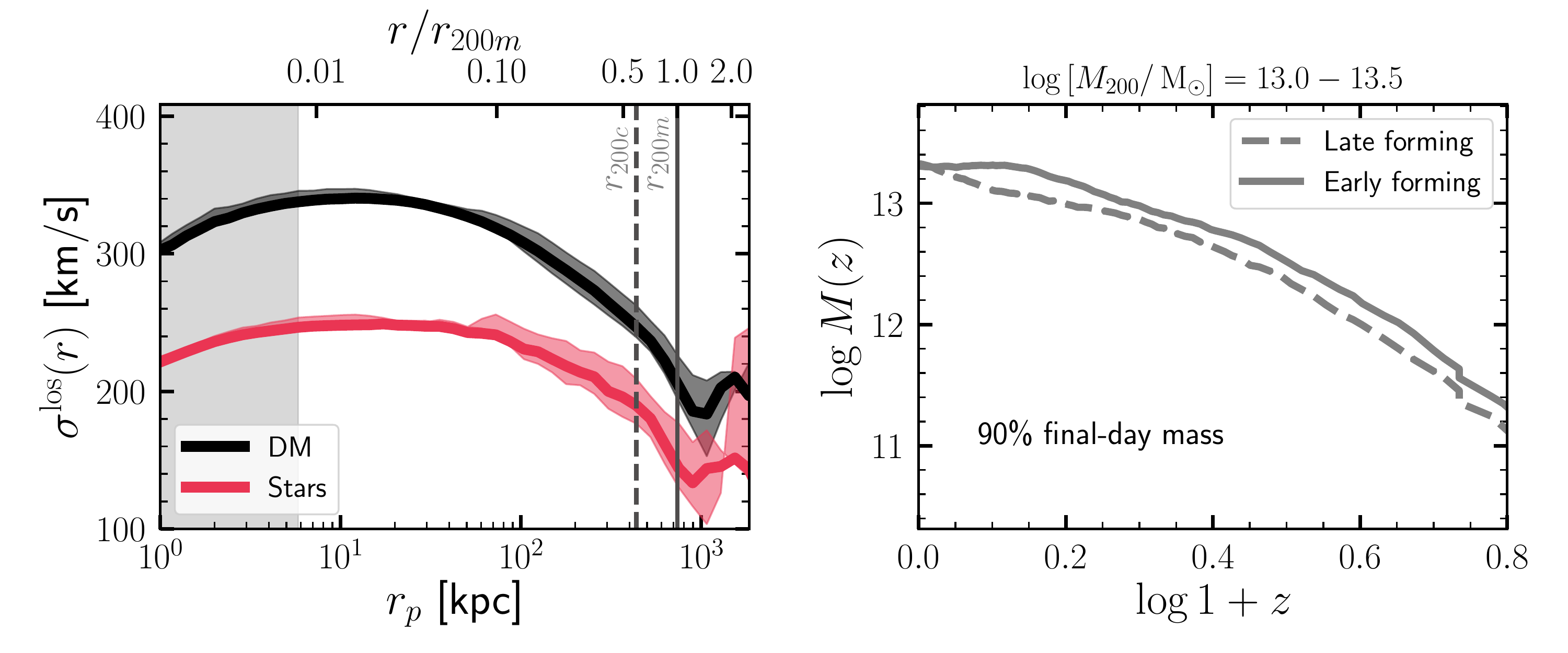} \\
\caption{The scatter in the line-of-sight velocity dispersion profiles of dark matter and stars after selecting halos by formation time {\it at fixed final-day mass} ($\log\left[M_{200}/{\rm M}_\odot\right]=13.0-13.5$ in this example). The profiles have been computed as a function of the projected halocentric radius, $r_p$. {\bf Top row}: selecting the 20\% earliest-forming and 20\% latest-forming halos as determined by the epoch by which 10\% of halo's $z=0$ dark matter content was assembled (right panel); the corresponding scatter in the line-of-sight velocity dispersion profiles, $\sigma^{{\rm los}}(r_p)$ is shown in the left panel as the shaded portions of the profiles. {\bf Bottom row}: as in the top row, but now selecting halos that are the 20\% earliest and latest-forming as defined by the redshift by which 90\% of their $z=0$ halo mass was assembled. This figure demonstrates that scatter in $\sigma^{{\rm los}}(r_p)$ in the inner parts of halos ($r \lesssim 0.1r_{200m}$) is driven primarily by scatter in the early assembly history of these halos; on the other hand, the scatter in $\sigma^{{\rm los}}(r_p)$ in the outskirts of halos is driven by variations in the late-time assembly of the same objects.}
\label{fig:los_disp_assembly}
\end{figure*}

In Figure~\ref{fig:los_disp_assembly}, we identify halos from TNG300 in the mass range $\log\left[ M_{200}/{\rm M}_\odot\right]=13.0-13.5$. We differentiate between halos with ``early'' and ``late-time'' formation histories by defining, respectively, the epoch at which 10\% (top row) and 90\% (bottom row) of the final-day (dark matter) mass of the halo has collapsed. For each definition, we then split halos that are the 20\% earliest-forming (solid lines) and 20\% latest-forming (dashed-lines). This selection therefore singles out halos at fixed $z=0$ mass that are most discrepant in their early and late-time formation histories. The left-hand panels in Figure~\ref{fig:los_disp_assembly} show the corresponding line-of-sight velocity dispersion profiles for dark matter and star particles. Solid lines show the mean stacked profile, while the shaded bands encompass the scatter around the mean, showing the range in profiles spanned by the halos that are picked out when selected according to their formation history.

From these two panels, we see a clear spatial dependence of the scatter in velocity dispersions when selecting on halo formation time. In particular, we find that halos that differ most in their late-time formation history show scatter in the outskirts of their velocity dispersion profiles ($r\gtrsim 0.1r_{200m}$). On the other hand, the scatter is larger in the regime of the inner profile when selecting halos that differ most in their early formation history.

\begin{figure*}
\centering
\includegraphics[width=0.475\textwidth]{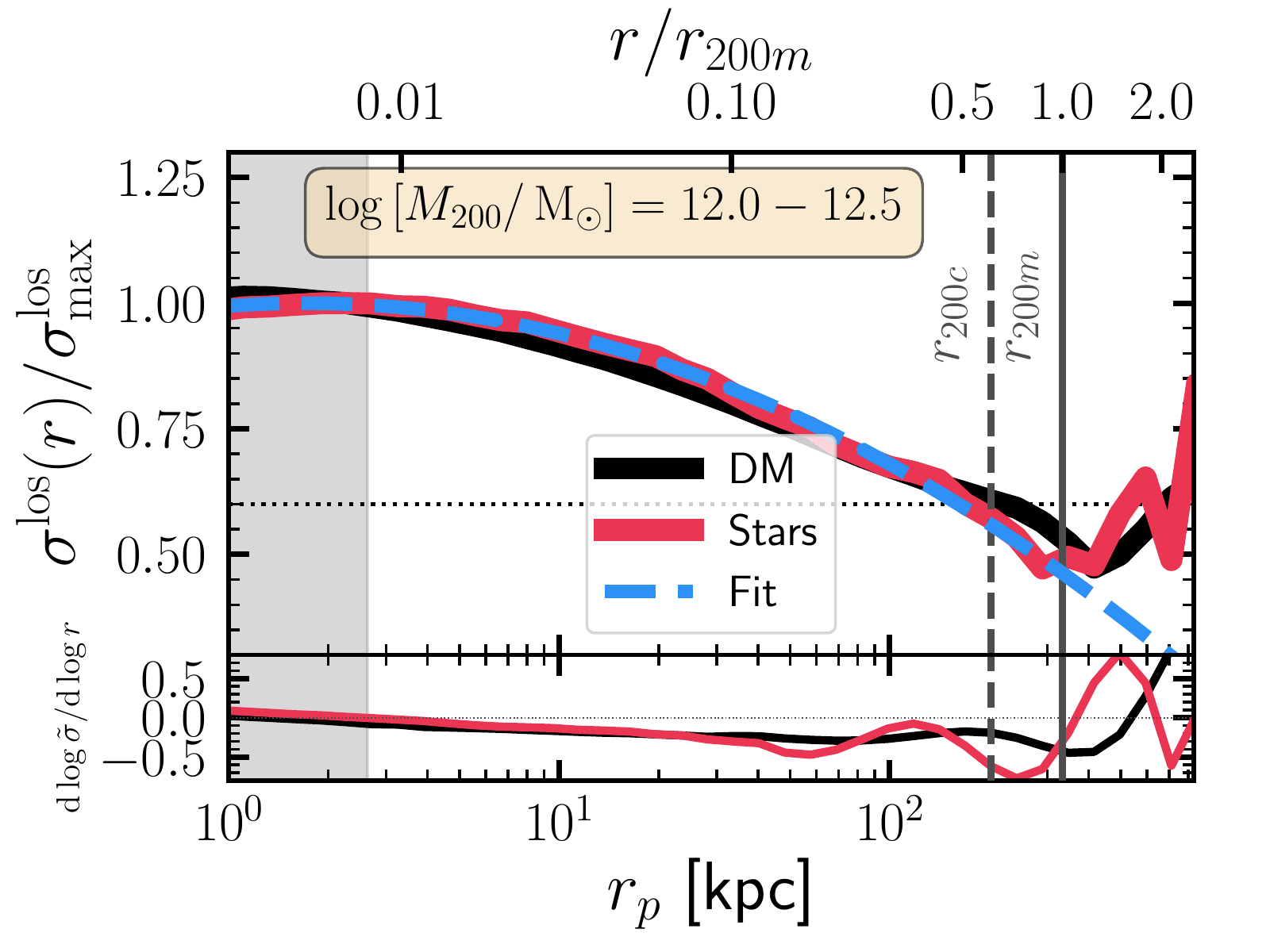}
\includegraphics[width=0.475\textwidth]{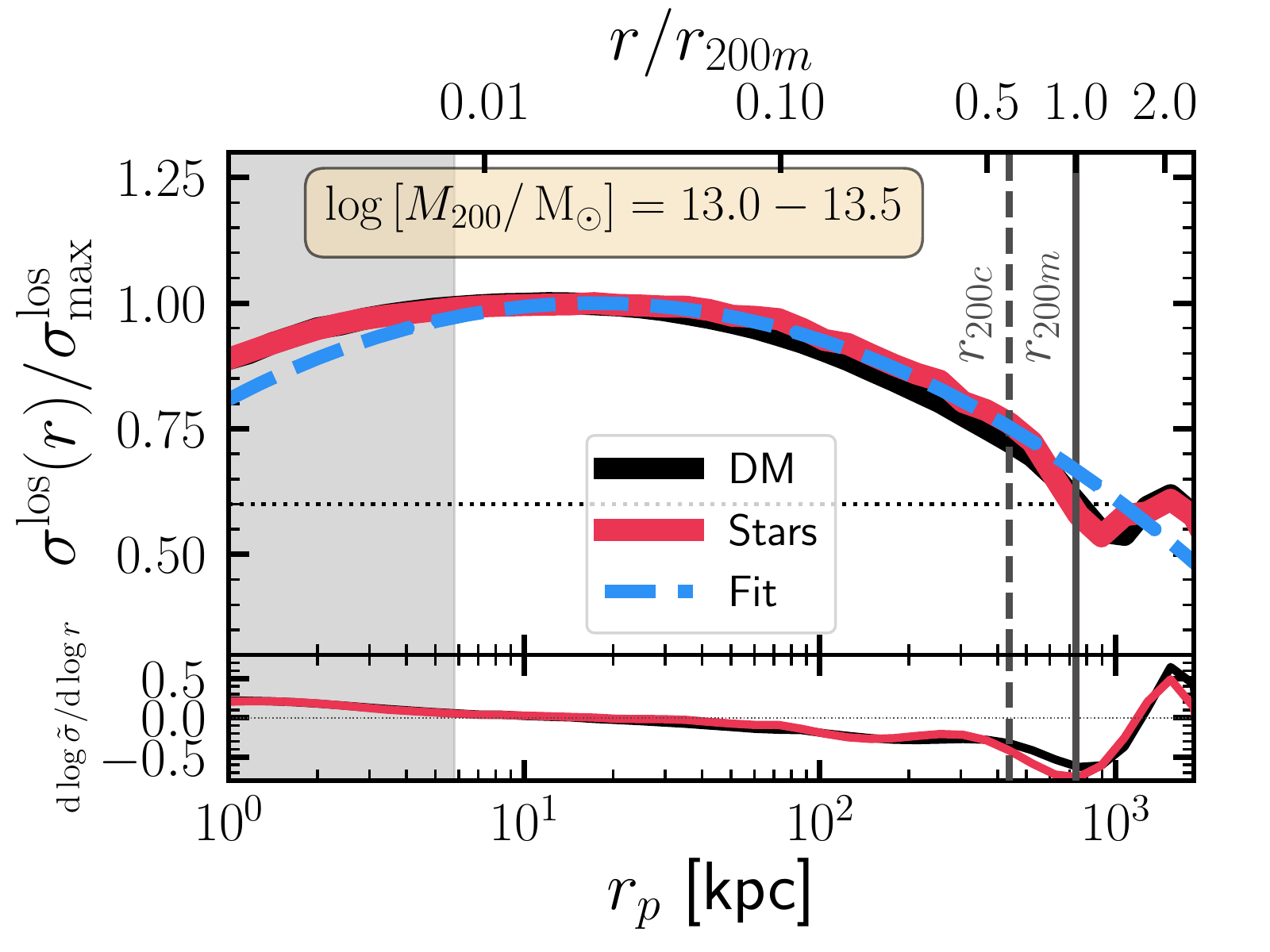}\\
\includegraphics[width=0.475\textwidth]{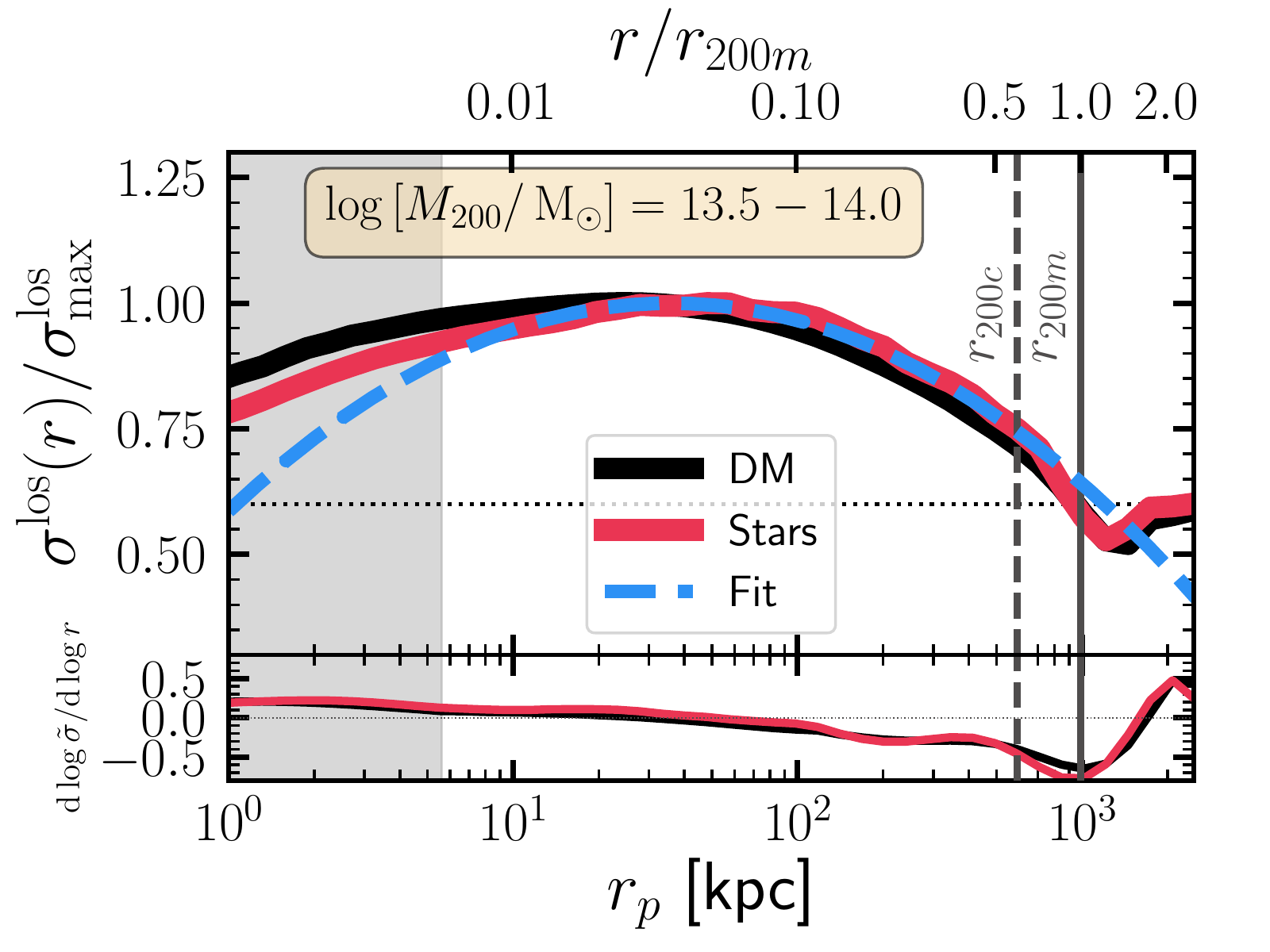}
\includegraphics[width=0.475\textwidth]{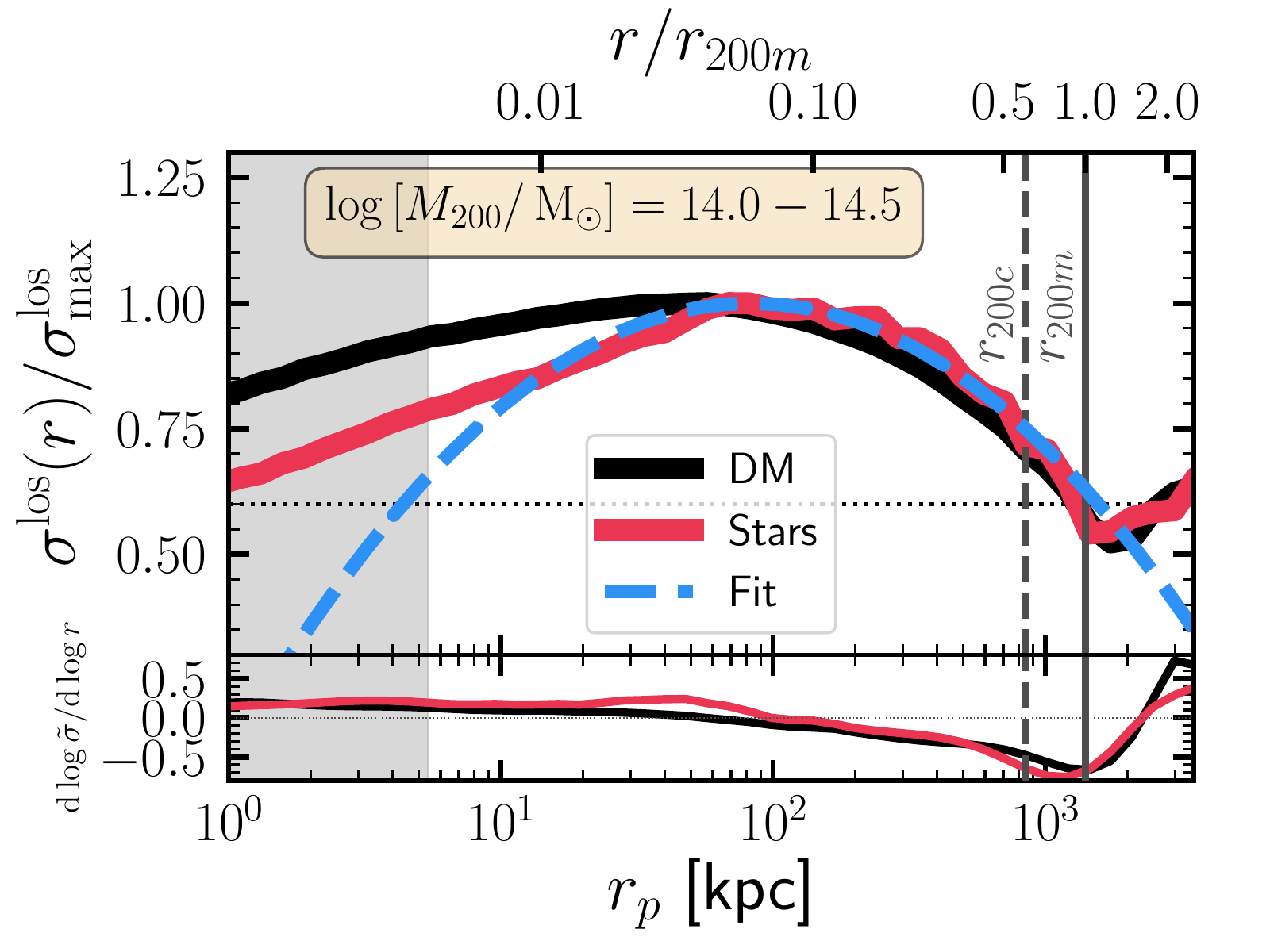}\\
\caption{Normalized line-of-sight velocity dispersion profiles for dark matter (black) and stars (red) as a function of projected halocentric radius, $r_p$, in TNG halos. The blue dashed curves represent fits to the stellar velocity dispersion profile using Eq.~(\ref{eq:disp_eq}). The lower sub-panels show the radial variation of the slope, ${\rm d} \log \tilde{\sigma} / {\rm d} \log r$, where $\tilde{\sigma} = \sigma^{{\rm los}}(r_p) / \sigma^{{\rm los}}_{{\rm max}}$. The horizontal dotted line marks the the case where the normalized stellar velocity dispersion falls to $60\%$ its maximum value; the intersection of this line with the velocity dispersion profile is typically consistent with $r_{200m}$. Note that this is also the location where the slope, ${\rm d}\log \tilde{\sigma} / {\rm d}\log r$, reaches its minimum value.}
\label{fig:los_disp_fit}
\end{figure*}

Figure~\ref{fig:los_disp_assembly} is an example of how the inside-out formation of dark matter halos manifests in the observable properties of visible tracers. In the standard picture of hierarchical structure formation, the central cores of halos collapse first, while the bulk of the mass of the halo continues to accumulate in the outskirts of halos through accretion and mergers. A tight scatter in velocity dispersions reflects regions of halos that are in some semblance of virial equilibrium. As the cores of halos collapse early on, selecting on differences in the early accretion histories of halos directly reflect differences in the central velocity dispersions of dark matter and stars. When selecting on differences in the late-time assembly history of halos, we specifically select objects that show large variances in the outskirts of their dispersion profiles. As seen in Figure~\ref{fig:los_disp_assembly}, the inner regions ($r\lesssim 0.1 r_{200m}$) show very little scatter as this portion of the halo is only marginally affected by late-time accretion events.

Figure~\ref{fig:los_disp_assembly} also shows that the impact of assembly history is imprinted in the dispersion profile for both dark matter and stars. The size of the scatter is also comparable between the two sets of tracers. The bottom row of this figure hints that scatter in the exterior profile may be somewhat larger for stars than for dark matter; this implies that following a late-time merger event, it takes longer for the stellar tracers to virialize than it does for the dark matter. The results suggest that luminous tracers may indeed be used to decipher aspects of the formation history of their host dark matter halos. In the following subsection, we investigate this in more detail and tie together the quantities that can be measured from the stellar velocity dispersion profiles with metrics that are associated with the host halo.

\subsection{A model for the velocity dispersion profiles of dark matter halos}
\label{sec:model}

In the previous subsection, we have seen a distinct connection between the early/late-time formation histories of dark matter halos and the imprint this leaves on the velocity dispersion profiles of dark matter and stars at present-day. In this subsection, we establish this relationship more formally.

The general shape of the normalized line-of-sight velocity dispersion profiles may, to first order, be described simply by a parabola, defined by a normalization term and the radial location of the peak of this function. While it is in principle possible to predict the dispersion profile starting from the NFW profile itself \citep[e.g.][]{Binney1982,Lokas2001}, here we define a simpler functional form and parametrize the {\it stellar} dispersion profile as:
\begin{equation} \label{eq:disp_eq}
\frac{\sigma^{{\rm los}}(r_p)}{\sigma^{{\rm los}}_{{\rm max}}} = 1 + \chi_\star \left( \log\left[ \frac{r_p}{r_\star}\right] \right)^2\;,
\end{equation}
in which $r_p$ is the projected distance from the halo center, $\chi_\star$ is a normalization factor, and $r_\star$ is the characteristic radius at which the profile is maximum ($\sigma^{{\rm los}}(r_\star) = \sigma^{{\rm los}}_{{\rm max}}$).

\begin{figure*}
    \centering
    \includegraphics[width=0.48\textwidth]{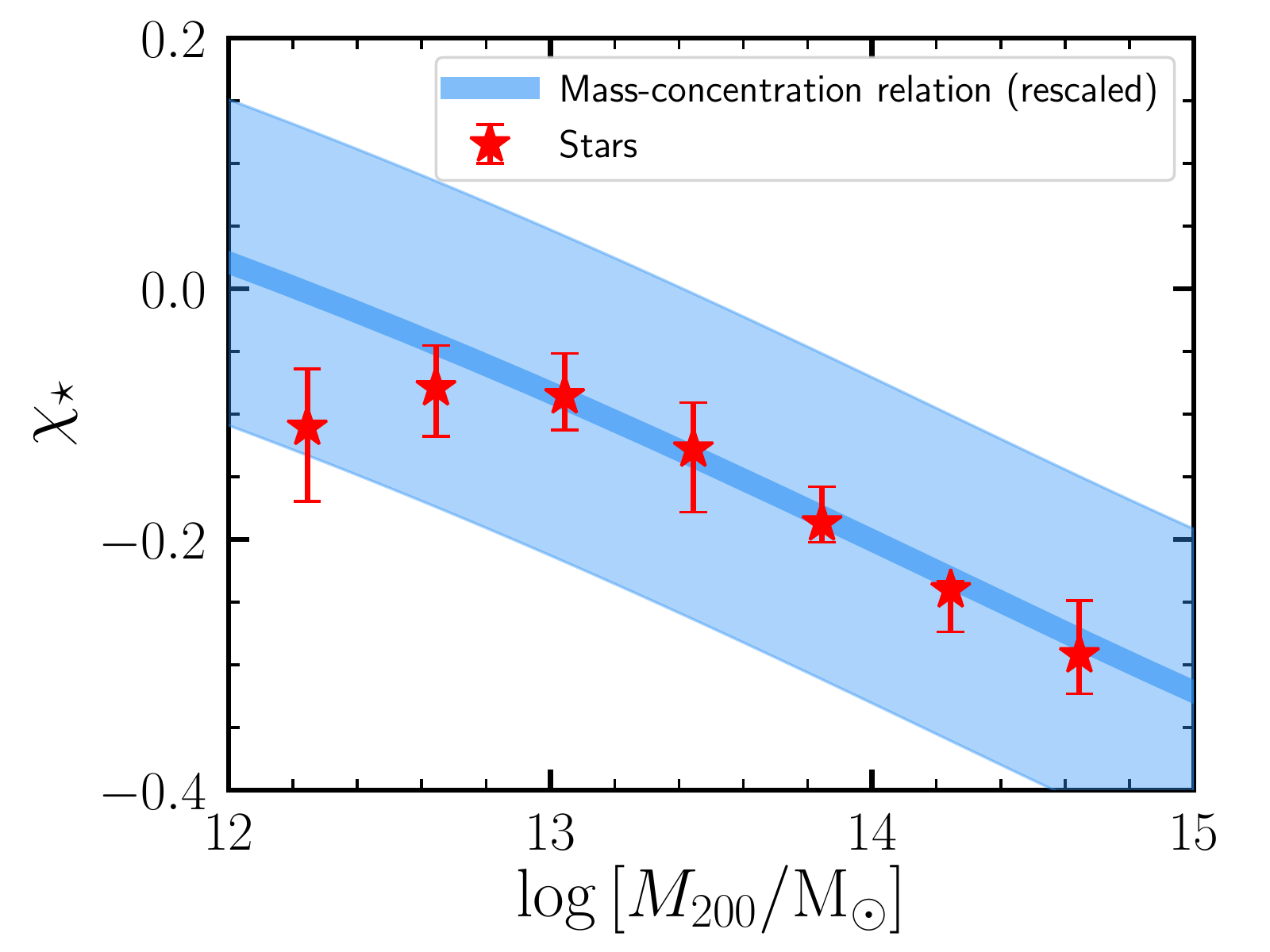}
    \includegraphics[width=0.48\textwidth]{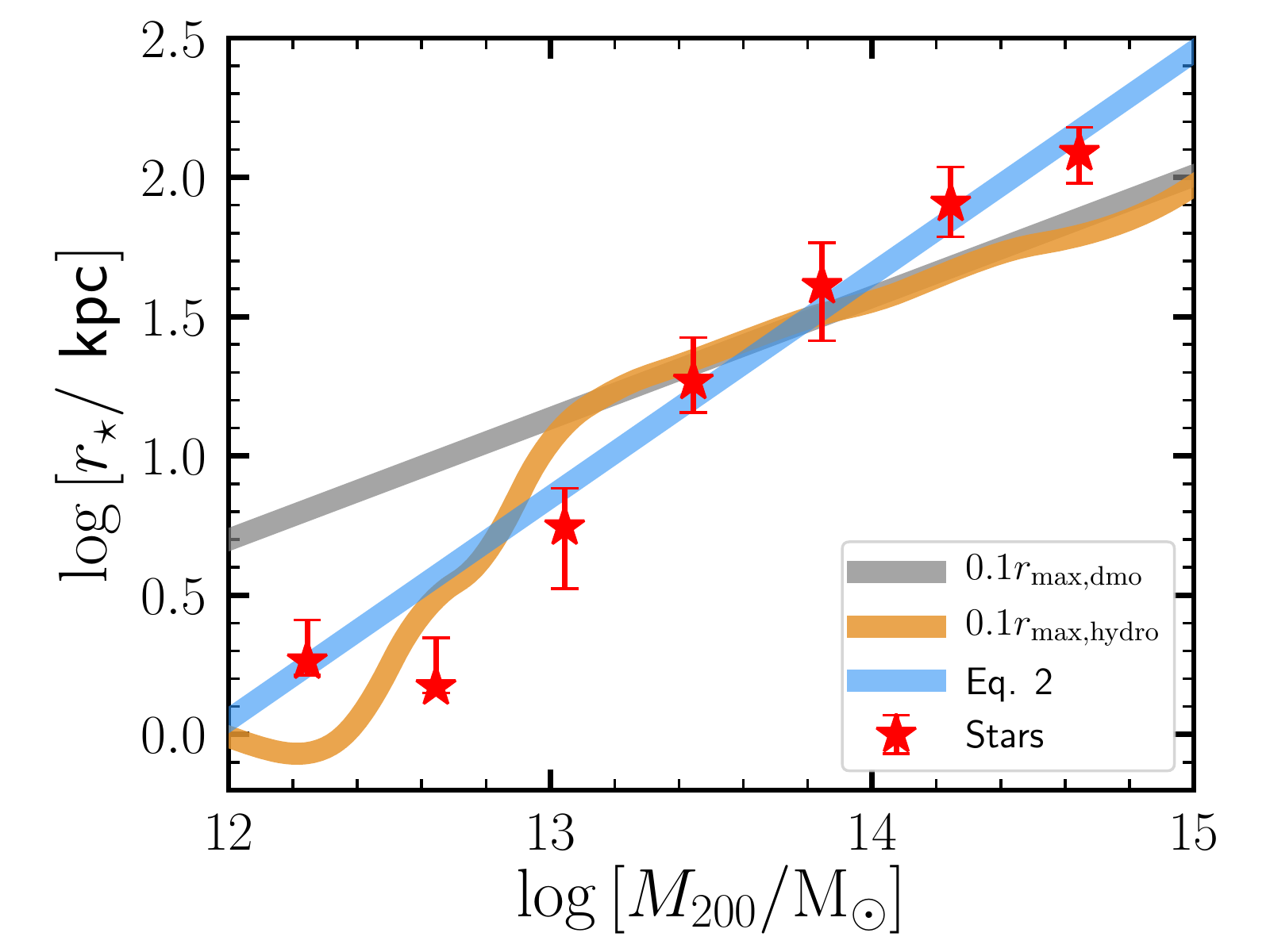}
    \caption{The halo mass dependence of the parameters $\chi_\star$ and $r_\star$ obtained through fitting Eq.~(\ref{eq:disp_eq}) to stellar velocity dispersion profiles in TNG. The stars represents the value of these parameters obtained from fitting the mean dispersion profile in any given mass bin, while the error bars show the range of $\chi_\star$ values measured from fitting the 16$^{\rm th}$-84$^{\rm th}$ percentile scatter around the mean profile. {\bf Left panel}: Comparison of the relationship between $M_{200}$ and $\chi_\star$ with the (rescaled) mass-concentration of DMO halos, as predicted by the model of \citet{Ludlow2016}. The shaded blue region encompasses the typical scatter in this relation of the order of 0.13 dex \citep[see, e.g.,][]{Dutton2014}. The halo mass dependence of $\chi_\star$ is very similar to that of the halo concentration, further establishing the intimate connection between the shape of the stellar velocity dispersion profiles and the assembly history of the host dark matter halo. The agreement worsens below $\log\left[ M_{200}/{\rm M}_\odot \right] \lesssim 12.5$, the mass scale below which the peak of dispersion profile is only marginally resolved in the TNG100 simulation (Figure~\ref{fig:los_disp_fit}). {\bf Right panel}: The halo mass dependence of $r_\star$, which is the characteristic radius at which the velocity dispersion profile reaches its peak value. The relationship between halo mass and $r_{{\rm max}}$, the radius at which the 3D circular velocity profile for TNG halos peaks is shown in orange (rescaled by a factor of 0.1 to aid comparison with $\chi_\star$). The same relation computed in the dark matter-only version of TNG is shown in gray. In all cases, we see a strong, positive correlation between halo mass and each of these characteristic radii. The blue line shows the best-fitting power law (defined in Eq.~\ref{eq:rstar_eq}) that describes the $M_{200}-r_\star$ relation measured in TNG.}
    \label{fig:mass_rmax_norm}
\end{figure*}

Figure~\ref{fig:los_disp_fit} shows the fits made to the normalized velocity dispersion profiles using Eq.~(\ref{eq:disp_eq}), separated in bins of host halo mass. The fit is performed only in the region to the right of the shaded gray box, which marks the radius within which particle dynamics are unreliable due to finite force resolution. The panels show only the mean normalized profiles, and we leave out the scatter for clarity. The blue dashed line is the best-fit profile obtained using Eq.~(\ref{eq:disp_eq}). The two parameter model provides a good fit to the dispersion profiles, across the full range of halo mass.

It is interesting to note that, when expressed in normalized units, the `kink' in the velocity dispersion profile -- which is related to the value of $r_{200m}$ (Section~\ref{sec:v3d}) -- appears at the radius at which $\sigma^{{\rm los}}(r_p) \approx 0.6\sigma^{{\rm los}}_{{\rm max}}$. This is certainly the case for the mean stacked profile in each mass bin, and the exact location depends on the assembly history of the halo (as indicated by the width of the shaded regions in Figure~\ref{fig:los_disp_assembly}). This suggests that it may be possible, to within a factor of a few, to {\it predict} the value of $r_{200m}$ by estimating the halocentric distance at which the line-of-sight velocity dispersion profile drops to $\sim 0.6$ its maximum value. We also note that at this radius, the logarithmic slope of normalized dispersion profile (denoted as $\tilde{\sigma}$) also reaches its minimum value; this is shown in the lower sub-panels in each frame in Figure~\ref{fig:los_disp_fit}.

The intersection of the horizontal dotted line in each panel of Figure~\ref{fig:los_disp_fit}, which marks the condition $\sigma^{{\rm los}}(r_p) \approx 0.6 \sigma^{{\rm los}}_{{\rm max}}$, with the velocity dispersion profile shows that indeed this radial location is coincident with $r_{200m}$. For objects in the range $\log\left[M_{200}/{\rm M}_\odot\right]>13.0$, the value of $r_{200m}$ estimated from the dispersion profile is quite close to its true value, while the agreement is substantially worse in the regime of Milky Way-mass halos (top-left panel of Figure~\ref{fig:los_disp_fit}). The reason for this may be partly due to the limited numerical resolution afforded by TNG100, the simulation box from which these halos have been extracted. As the top-left panel shows, the full parabolic shape is not resolved fully in TNG100 for this mass range. In particular, profile peaks below the convergence radius, the portion of the halo internal to which issues pertaining to finite force softening start to dominate (gray shaded region).

It is worthwhile exploring the physical interpretation of the fit parameters $\chi_\star$ and $r_\star$ in more detail -- in particular, their connection to the properties of the host dark matter halo. Figure~\ref{fig:mass_rmax_norm} shows the mass dependence of these parameters as obtained through fitting Eq.~(\ref{eq:disp_eq}) to the velocity dispersion profiles of the stars in TNG halos.

\begin{figure*}
    \centering
    \includegraphics[width=\textwidth]{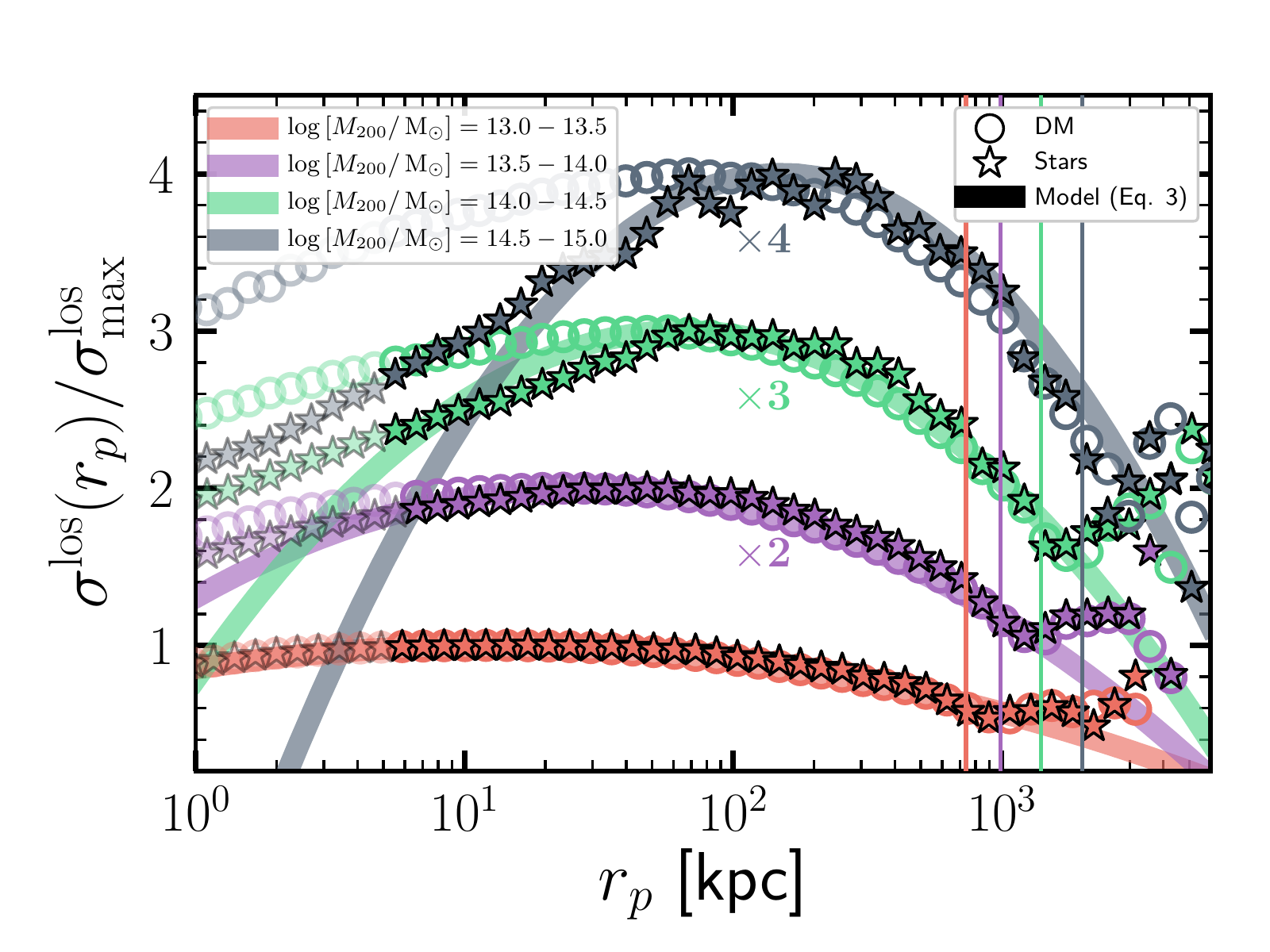}
    \caption{Normalized line-of-sight velocity dispersion profiles for dark matter and stars as a function of projected halocentric radius, $r_p$, in TNG halos. The colors represent results from different mass bins, and each pair of profiles has been offset vertically for clarity. Furthermore, the symbols representing individual profiles are made fainter below the convergence radius. The vertical solid lines mark the location of $r_{200m}$ for the corresponding mass bin. The thick curves represent predictions of the halo mass and concentration-based model defined in Eq.~(\ref{eq:disp_eq_recast}), where the concentration for any given mass bin is predicted using the \citet{Ludlow2016} model. We find that across the range of halo masses considered in this work, the model in Eq.~(\ref{eq:disp_eq_recast}) provides a good fit to the velocity dispersion profiles measured in the TNG simulations.}
    \label{fig:model_compare}
\end{figure*}

\begin{figure}
    \centering
    \includegraphics[width=\columnwidth]{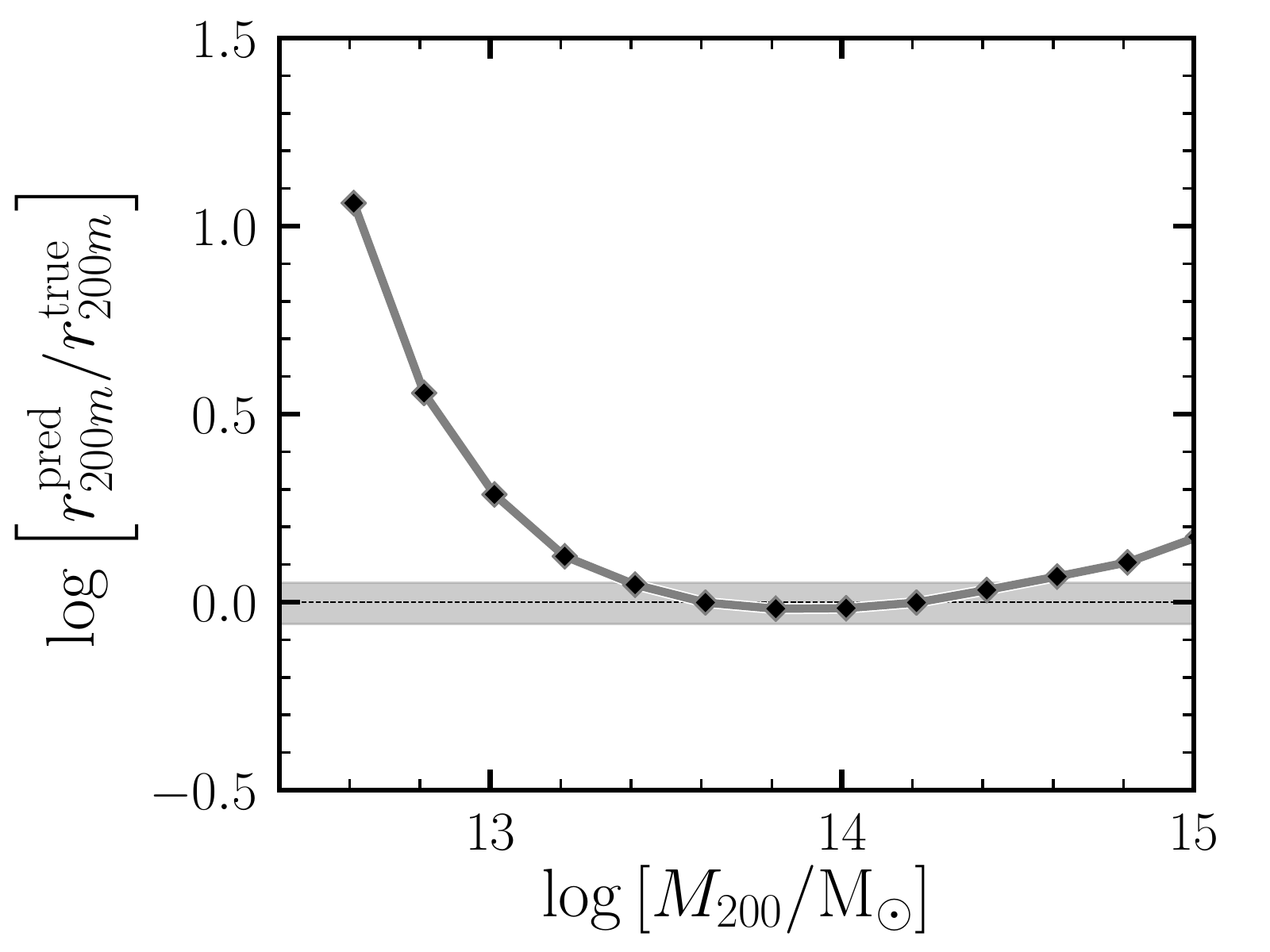}
    \caption{Ratio between the true and predicted values of $r_{200m}$,  where  the  latter is  determined  by  using  Eq.~(\ref{eq:disp_eq_recast}) to estimate where the normalized line-of-sight velocity dispersion profile drops to $\sim 0.6$ its peak value (see main text for details).  The gray shaded band marks the 25\% error region around the true value. For halos more massive than low-mass groups,  which are the best-resolved objects in our dataset,the agreement is generally quite good, often to within 25\%.The level of agreement towards lower halo mass is substantially poorer.}
    \label{fig:rvir_pred}
\end{figure}

\begin{figure}
    \centering
    \includegraphics[width=\columnwidth]{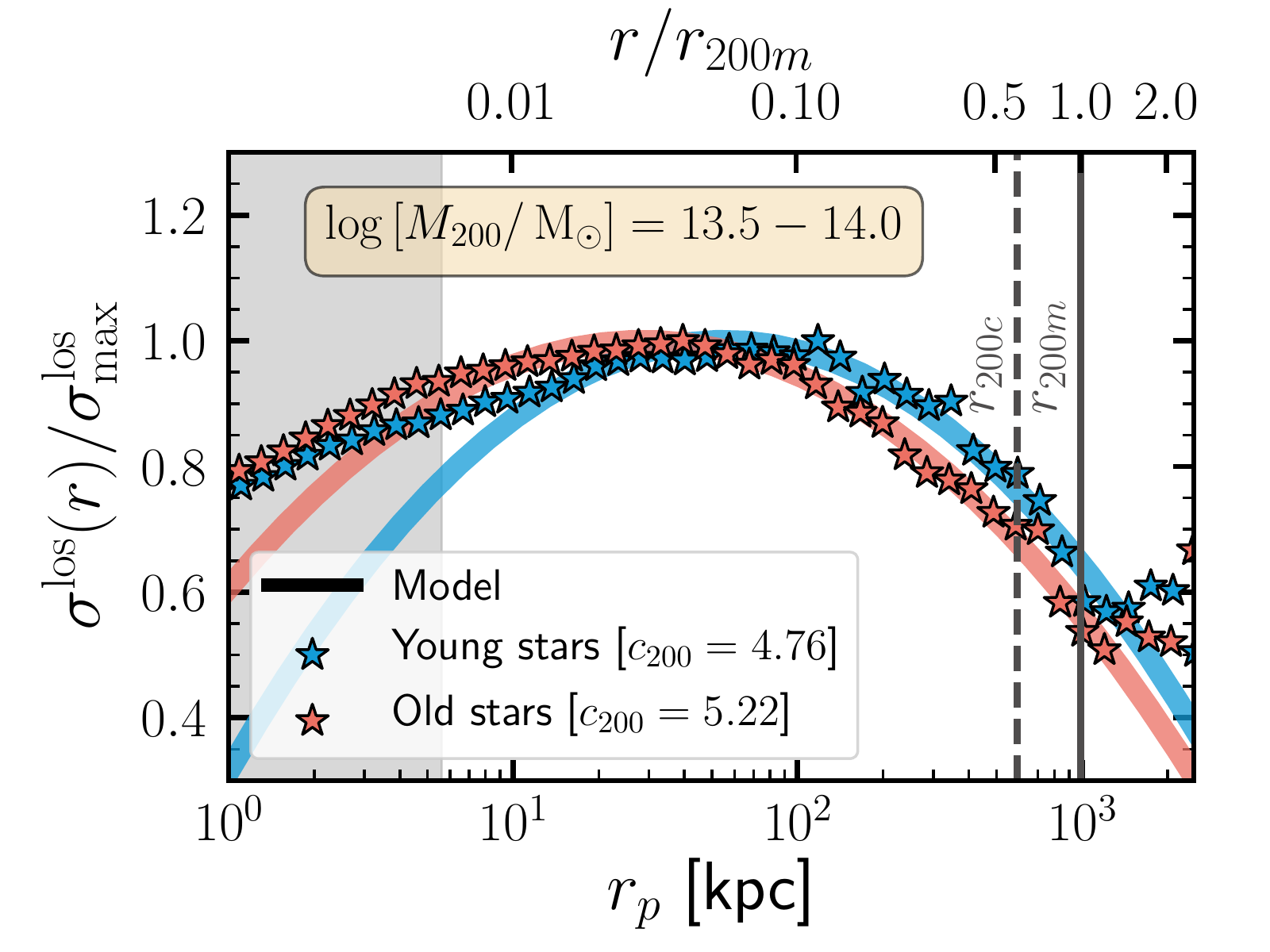}
    \caption{The normalized line-of-sight velocity dispersion profile of halos in the mass range $\log\left[M_{200}/{\rm M}_\odot\right]=13.5-14.0$, split by {\it stellar age}. The  objects are split into two halves: those with `older' stars, represented in red, and those with `younger' stars shown in blue. Here, stellar age is defined as the epoch at which the halo accumulated 50\% of its final-day {\it stellar} mass. The measurements from the simulated halos are represented by the star symbols; the smooth curves are fits to the measured profiles obtained using Eq.~(\ref{eq:disp_eq_recast}). As noted in the legend, we find that our model successfully predicts the expected trend where older objects have higher concentrations ($c_{200}$) than younger ones, although the difference within a fixed mass bin is small.}
    \label{fig:vlos_stellar_fit}
\end{figure}

Mathematically, the normalization parameter, $\chi_\star$, determines the concavity (or width) of the profile which, as we discussed in Section~\ref{sec:assembly}, is related to the assembly history of the host halo. For dark matter halos, a simple parameter that is often used to characterize the assembly history of halos is the concentration, $c_{200}$, defined as the ratio $r_{200c}/r_s$ where, for a dark matter halo described by an NFW profile, $r_s$ is the `scale radius' at which the slope of the density profile is $-2$.

The solid blue line in the left-hand panel of Figure~\ref{fig:mass_rmax_norm} shows the median halo mass-concentration relation for dark matter halos in the {\it Planck} cosmology, as predicted by the model described in \cite{Ludlow2016}. In this model, the concentration of NFW halos may be predicted using only the initial power spectrum, which determines (at least to first order) the subsequent assembly history of the halo. The predictions of the \cite{Ludlow2016} model have been rescaled to facilitate comparison of their relationship with $\chi_\star$. The blue shaded region shows a constant scatter of 0.13 dex in the halo mass-concentration relation \citep[see, e.g.,][]{Dutton2014}. 

It is striking to note that the gradient of the $M_{200}$-$\chi_\star$ relation obtained from the stellar velocity dispersion profile is nearly identical to that of the $M_{200}$-$c_{200}$ of DMO halos across this mass scale (albeit with a different normalization). This reaffirms our earlier interpretation of $\chi_\star$ as a parameterization of the assembly history of the host dark matter halo. This suggests that given a halo of mass $M_{200}$, the parameter $\chi_\star$, which sets the normalization of the profile in Eq.~(\ref{eq:disp_eq}), may be {\it predicted} using the standard mass-concentration relation of dark matter halos. 

Focusing next on the relationship between $r_\star$ and $M_{200}$ (right-hand panel of Figure~\ref{fig:mass_rmax_norm}), we find a strong correlation between these quantities. This is unsurprising: the radius at which the velocity dispersion profile peaks is tied directly to the depth of the potential well of the halo (and, therefore, its mass). Indeed, the parameter $r_\star$ bears a similar relationship with halo mass as the more familiar parameter, $r_{{\rm max}}$, which denotes the radius at which the halo's 3D circular velocity profile reaches its maximum. This is represented by the solid gray line (rescaled by a factor of 0.1), which shows the $M_{200}$-$r_{{\rm max}}$ relationship for halos in this mass range in TNG DMO; the corresponding relation for the full physics TNG simulation is shown in orange. 

The $M_{200}-r_{{\rm max}}$ relation, which is nearly identical in the full physics and DMO versions of TNG in the mass scale $\log\left[M_{200}/{\rm M}_\odot\right] \gtrsim 13.0$, steepens at lower masses in the hydrodynamical simulation. This is likely the consequence of gas cooling and subsequent star formation, which is concentrated primarily in the central regions of halos, thereby contracting the radius at which the peak circular velocity is achieved in the hydrodynamical simulation compared to TNG DMO; the effect on the largest halos is modest. Both observations are consistent with the conclusions of \cite{Lovell2018}, who investigated the impact of baryon physics on the dark matter content of TNG halos. The $M_{200}$-$r_\star$ relationship measured from the stellar velocity dispersion profile shows a similar gradient to the $M_{200}$-$r_{{\rm max,hydro}}$ in the mass scale $\log\left[M_{200}/{\rm M}_\odot\right] \lesssim 14.0$, but exhibits a somewhat steeper mass dependence in the regime of galaxy clusters.

In TNG, we find that the relationship between $r_\star$ and $M_{200}$ can be described by the following functional form:
\begin{equation} \label{eq:rstar_eq}
    r_\star = 7.1 \left( \frac{M_{200}}{10^{13}\,{\rm M}_\odot} \right)^{0.8} {\rm kpc};
\end{equation}
which, when combined with the realization that the normalization parameter $\chi_\star$ may be predicted by the mass-concentration relation allows us to recast Eq.~(\ref{eq:disp_eq}) in terms of halo mass, $M_{200}$, and halo concentration, $c_{200}$, only:
\begin{equation} \label{eq:disp_eq_recast}
    \frac{\sigma^{{\rm los}}(r_p)}{\sigma^{{\rm los}}_{{\rm max}}} = 1 + \log \left(\frac{c_{200}}{8}\right)    \log^2 \left[ \frac{r_p}{7.1 \,{\rm kpc}} \left(\frac{10^{13}\,{\rm M}_\odot}{M_{200}}\right)^{0.8}\right].
\end{equation}
The fact that line-of-sight velocity dispersion profile can be expressed in terms of halo mass and concentration should come as no surprise since it is indeed possible to predict the velocity dispersion starting from the NFW profile itself \citep[see, e.g.][for the functional form of the {\it radial} velocity dispersion in terms of the same quantities]{More2009}. The formula expressed in Eq.~(\ref{eq:disp_eq_recast}) provides a simple and convenient functional form that accurately captures the form of these profiles measured in TNG. 

In Figure~\ref{fig:model_compare}, we show comparisons between predictions of the model defined by Eq.~(\ref{eq:disp_eq_recast}) (solid lines) and the dispersion profiles actually measured in our simulations (symbols, made fainter below the convergence radius). The open circles represent the dispersion profiles of the dark matter component, while the filled stars represent the stellar velocity dispersion profile. Different bins of halo mass are represented by the different colors. Lines and symbols of a given color have been renormalized by a constant factor to offset them for clarity. The average location of $r_{200m}$ is shown using the vertical solid lines. 

The model defined in Eq.~(\ref{eq:disp_eq_recast}), in which the only two free parameters are the mass and concentration of the halo, does a good job of describing the shape of the velocity dispersion profiles across the full range of host halo masses. Given the relatively simple parameterization that we have adopted, the model fails to fully capture the intricacies of the dispersion profile, particularly at its extremities, but the quality of the fit is good in and around the peak of the profile. Using the fact that the tentative location of $r_{200m}$ may be identified as the radius at which velocity dispersion reaches $\sim0.6$ its peak value, we can extrapolate the smooth profiles defined by Eq.~(\ref{eq:disp_eq_recast}) to estimate this radius in each mass bin. 

Figure~\ref{fig:rvir_pred} compares the quality of our predictions for $r_{200m}$ (using Eq.~\ref{eq:disp_eq_recast}) quantitatively. Here, we show the ratio of the predicted value to the true value of $r_{200m}$ as a function of halo mass. The gray shaded band shows a 25\% error region about the true value. In general, we find that the value of $r_{200m}$ estimated by extrapolating Eq.~(\ref{eq:disp_eq}) to $\sigma^{{\rm los}}(r_p) \approx 0.6 \sigma^{{\rm los}}_{{\rm max}}$ agrees with the true, averaged value of $r_{200m}$ to within a factor of two, and often to within 25\%. The exceptions are at the extreme ends of the mass scale, where finite resolution affects our measurements at the scale of halos less massive than groups, while finite box size (i.e. statistics) in TNG300 affects our ability to measure the profile at the scale of rich clusters. The behavior of this ratio at the low mass end stresses the importance of resolving the peak of the dispersion profile when using Eq.~(\ref{eq:disp_eq_recast}) to predict $r_{200m}$. Note that the exercise of measuring the value of $r_{200m}$ by estimating where the dispersion profile falls to 60\% of its peak value is most effective when the dispersion profiles of several objects have been stacked. Individual galaxies are likely to be noisy and will exhibit significant deviation from spherical symmetry (see Figure~\ref{fig:3d_disp}); the stacking procedure helps in reducing this noise. Stacking will, of course, also be necessary to probe the exterior portion of the dispersion profile, where the surface brightness in individual galaxies is extremely low.

It is an instructive exercise to compare the quality of these predictions with  existing observational techniques for measuring the boundary of halos. Recent efforts include the measurement of the weak lensing mass profile \citep{Chang2018} or of (projected) galaxy density profiles around massive clusters (typically with mass $ \log \left[M_{200}/{\rm M}_\odot \right] \gtrsim 14.2$), selected either optically \citep[e.g.][]{More2016,Baxter2017,Murata2020,Bianconi2020} or via the Sunyaev-Zel'dovich effect \citep{Shin2019}. These techniques have typically measured the splashback radius to within 10-20\% of the values predicted from $N$-body simulations. Yet another way to estimate the virial radius (or, more accurately, the virial mass) of halos from observations is to use the halo abundance matching technique, which assigns the measured stellar mass to halo mass by pairing galaxies with halos with the same cumulative number densities \citep[e.g.][]{Kravtsov2004,Tasitsiomi2004,Vale2004}. After applying the abundance matching to technique to a sample of galaxies from the Sloan Digital Sky Survey, \cite{Calderon2019} find this method to yield errors in the predicted halo mass at the 0.27-0.90 dex level; the error depends on the mass scale of interest, as well as the halo property used as a proxy for abundance matching (halo mass, maximum circular velocity etc.). In comparison to these methods, the methodology for using the velocity dispersion profile for estimating the halo virial radius as outlined in this paper is competitive, particularly in the regime of clusters. Furthermore, it serves as an orthogonal measurement, which can be employed jointly with existing techniques to obtain more robust estimates of the halo virial radius.

As a final test of our main conclusions, in Figure~\ref{fig:vlos_stellar_fit} we show the normalized line-of-sight velocity dispersion profiles of well-resolved halos in the mass range $\log\left[M_{200}/{\rm M}_\odot\right]=13.5-14.0$, split now by the ages of their {\it stellar populations}. We use a definition analogous to the one for halo formation time, such that the stellar age of a halo is defined as the epoch at which 50\% of the object's final-day {\it stellar mass} was accumulated. The 50\% oldest halos according to this definition are shown in red, while the remaining 50\% younger halos are shown in blue.

It is immediately clear that selecting halos by stellar age (at fixed mass) also separates them in the space of the velocity dispersion profiles, just as it did when selecting on halo formation time (Figure~\ref{fig:los_disp_assembly}). The solid lines are the best-fit profiles obtained using the model in Eq.~(\ref{eq:disp_eq_recast}); the best-fit concentration, $c_{200}$, is listed in the legend. We find that our model predicts that older halos have higher concentrations than younger halos, as expected. Furthermore, the range of concentrations predicted is also consistent with the values expected of halos at this mass scale. The difference in concentration between the old and young halo subsets is not large ($\sim 10\%$). Indeed, the predicted difference is larger if the populations are selected using {\it halo} formation time instead, which is to be expected given that the concentration is tied more directly to the assembly of the dark matter halo than it is the assembly of the stars. Finally, we also note that the familiar kink in the outskirts of the profile are present in both the young and old stellar population subsets.

The results presented in this section suggest that the velocity dispersion profiles of stars contain substantial amounts of information on the dark matter halos in which they reside. In particular, they encode both the mass and the memories of the accretion history of the halo. The function presented in Eq.~(\ref{eq:disp_eq}) contains two free parameters, $r_\star$ and $\chi_\star$ which capture this information. Furthermore, these quantities are closely associated with their analogues in NFW halos formed in DMO simulations, $r_{{\rm max}}$ and the concentration, $c_{200}$, which allows us to then rewrite Eq.~(\ref{eq:disp_eq}) into a form that depends on halo mass and concentration explicitly (Eq.~\ref{eq:disp_eq_recast}). In principle, therefore, a measurement of the stellar velocity dispersion profile allows a potential avenue to infer halo mass and concentration directly.

\section{Conclusions}
\label{sec:conclusions}

In this paper, we have explored the dynamics of dark and luminous tracers of the potential wells of dark matter halos. In particular, we used the \tng{} cosmological, hydrodynamical simulations to measure the (line-of-sight) velocity dispersion profiles of dark matter and stars, $\sigma^{{\rm los}}(r_p)$, and established the connection between the form of these profiles and the properties of the encompassing dark matter halo. In order to sample a wide range of halo mass, we combined data from the publicly-available 100 Mpc (TNG100) and 300 Mpc (TNG300) TNG simulation volumes. Our main results are summarized as follows:

\begin{itemize}

\item The velocity dispersion profiles for halos exhibit a universal shape (as a function of halo mass), and is nearly identical for both dark matter and stars. The exterior profile shows a characteristic `kink', on average around $\sim 1.2\,r_{200m}$, consistent with the expected location of the so-called `splashback radius' of the halo (Figure~\ref{fig:3d_disp}).

\item Halos at fixed mass exhibit significant scatter around the mean dispersion profile shape. This scatter may be explained, at least in part, by differences in the assembly history of the halo compared to average assembly history of halos in that mass bin. In particular, the late-time formation history of the halo influences the shape of the exterior portion of the velocity dispersion profile ($r\gtrsim 0.1 r_{200m}$), while the region interior to this is more strongly affected by the early accretion history of the halo (Figure~\ref{fig:los_disp_assembly}).

\item When expressed in normalized units, the kink in the exterior of the profile occurs approximately where the velocity dispersion profile drops to 60\% its peak value ($\sigma^{{\rm los}}(r_p) \approx 0.6 \sigma^{{\rm los}}_{{\rm max}}$). The (normalized) velocity dispersion profiles obtained from both dark  matter and stars are well fit by the two parameter functional form presented in Eq.~(\ref{eq:disp_eq}). This simple model provides a good fit to the velocity dispersion profile across a wide range of halo mass (Figure~\ref{fig:los_disp_fit}).

\item There are distinct connections between the fit parameters in Eq.~(\ref{eq:disp_eq}) and the properties of the host dark matter halo (Figure~\ref{fig:mass_rmax_norm}). In particular, we find that the coefficient in this equation, $\chi_\star$, bears a near identical dependence on halo mass as the concentration of the host halo. The characteristic length scale in this fitting formula, $r_\star$, is directly correlated with the mass of the halo (Eq.~\ref{eq:rstar_eq}).

\item The consequence of this relationship is that it allows us to recast Eq.~(\ref{eq:disp_eq}) in terms of halo mass and concentration only, in the form described in Eq.~(\ref{eq:disp_eq_recast}). The resulting model does a reasonably good job of matching the shape of the velocity dispersion profiles measured in the TNG simulation suite, across the full range of halo masses considered in this work (Figure~\ref{fig:model_compare}). This figure also suggests that halo mass and concentration may be inferred directly from (stacked) stellar velocity dispersion profiles measured in galaxies. 

\item Using Eq.~(\ref{eq:disp_eq_recast}) to estimate where $\sigma^{{\rm los}}(r_p) \approx 0.6 \sigma^{{\rm los}}_{{\rm max}}$, one can `predict' the virial radius, $r_{200m}$, based on the measured velocity dispersion profile. In general, the predicted and true values of $r_{200m}$ agree to within a factor of two, although it is worse for halos with mass $\log\left[M_{200}/{\rm M}_\odot\right]<13.0$, where we do not properly resolve the peak of the dispersion profile (Figure~\ref{fig:rvir_pred}).

\item The velocity dispersion profiles also reflect differences in the assembly of the stellar component of halos at fixed mass. In particular, we find clear differences in the velocity dispersion profiles of halos selected based on young/old stellar populations, in a manner that is consistent with the dependence on the halo formation time (Figure~\ref{fig:vlos_stellar_fit}). Encouragingly, we find that using Eq.~(\ref{eq:disp_eq_recast}) to estimate the concentrations of halos at fixed mass, split by stellar age, predicts that halos with younger stellar populations have lower concentrations on average than halos with an older stellar component, as expected.

\end{itemize}

Our study demonstrates the value in measuring the dynamics of tracers of the gravitational potential in the very outskirts of galaxies. Accurate measurements in this regime may inform us of both the mass and details of the assembly history of the host dark matter halo. Indeed, the measurement of the outskirts of these entities may soon be within reach, at least after stacking profiles of several objects. For example, the surface brightness of massive clusters around $r_{200m}$ ranges between $32-36$ mag arcsec$^{-2}$ at $z=0.25$ \citep[e.g.][]{Deason2020b}, which will be within the operational capabilities of future observational facilities like the Rubin Observatory Legacy Survey of Space and Time \citep[LSST,][]{Ivezic2019}, {\it Euclid} \citep{Laureijs2011}, and {\it The Nancy Grace Roman Space Telescope} \citep{Spergel2015}. As these and other facilities geared towards low surface brightness measurements begin to come online \citep[see][for a review]{Kaviraj2020}, the feasibility of running the deep and wide surveys necessary for this kind of measurement will become reality.

\acknowledgments

We are grateful to the referee for providing us with a constructive report, and for suggesting the test in Figure~\ref{fig:vlos_stellar_fit}, enhancing the scope of our results and the overall quality of this work. We thank the IllustrisTNG collaboration for making the data used in this paper available for public use (\href{https://www.tng-project.org/}{https://www.tng-project.org/}). We thank Benedikt Diemer for insightful discussions during the course of this project, and for providing helpful feedback on this manuscript. We are also grateful to Lars Hernquist and Ken Freeman for their input at the onset of this project. This project made use of the SAO/NASA Astrophysics Data System (ADS), the arXiv.org preprint server, as well as the {\tt matplotlib} \citep{Hunter2007}, {\tt numpy} \citep{Numpy2020}, and {\tt scipy} \citep{Scipy2020} python packages. S.B. is supported by Harvard University through the ITC Fellowship.

\appendix
\section{Numerical convergence}
\label{sec:convergence}

In this section, we show the level of convergence in the velocity dispersion profiles measured in the TNG simulations. By comparing the stacked profiles across two levels of numerical resolution, we are able to diagnose the extent to which the features we have identified in Section~\ref{sec:results} are ``real'', as opposed to being numerical artifacts. To this end, we compare results from two versions of the TNG300 simulation: TNG300-1, which is the high-resolution, fiducial dataset used in this work, and TNG300-2, its low-resolution counterpart. With $2\times1250^3$ resolution elements in the same 300 Mpc box, the TNG300-2 simulation has an effective mass resolution that is $8\times$ poorer than TNG300-1, but is run using the same initial phases and the same underlying galaxy formation model. 

\begin{figure*}
    \centering
    \includegraphics[width=0.475\textwidth]{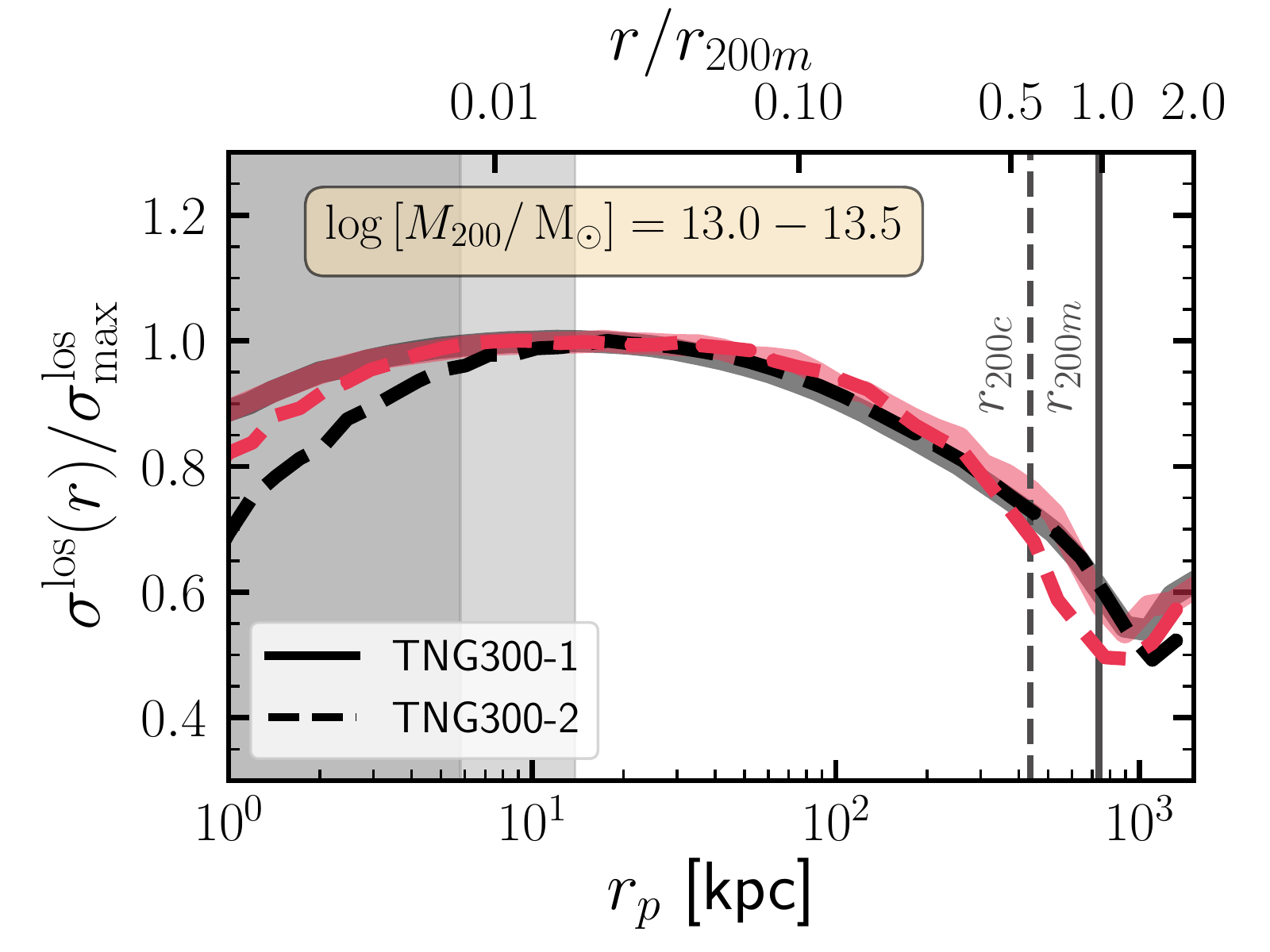}
    \includegraphics[width=0.475\textwidth]{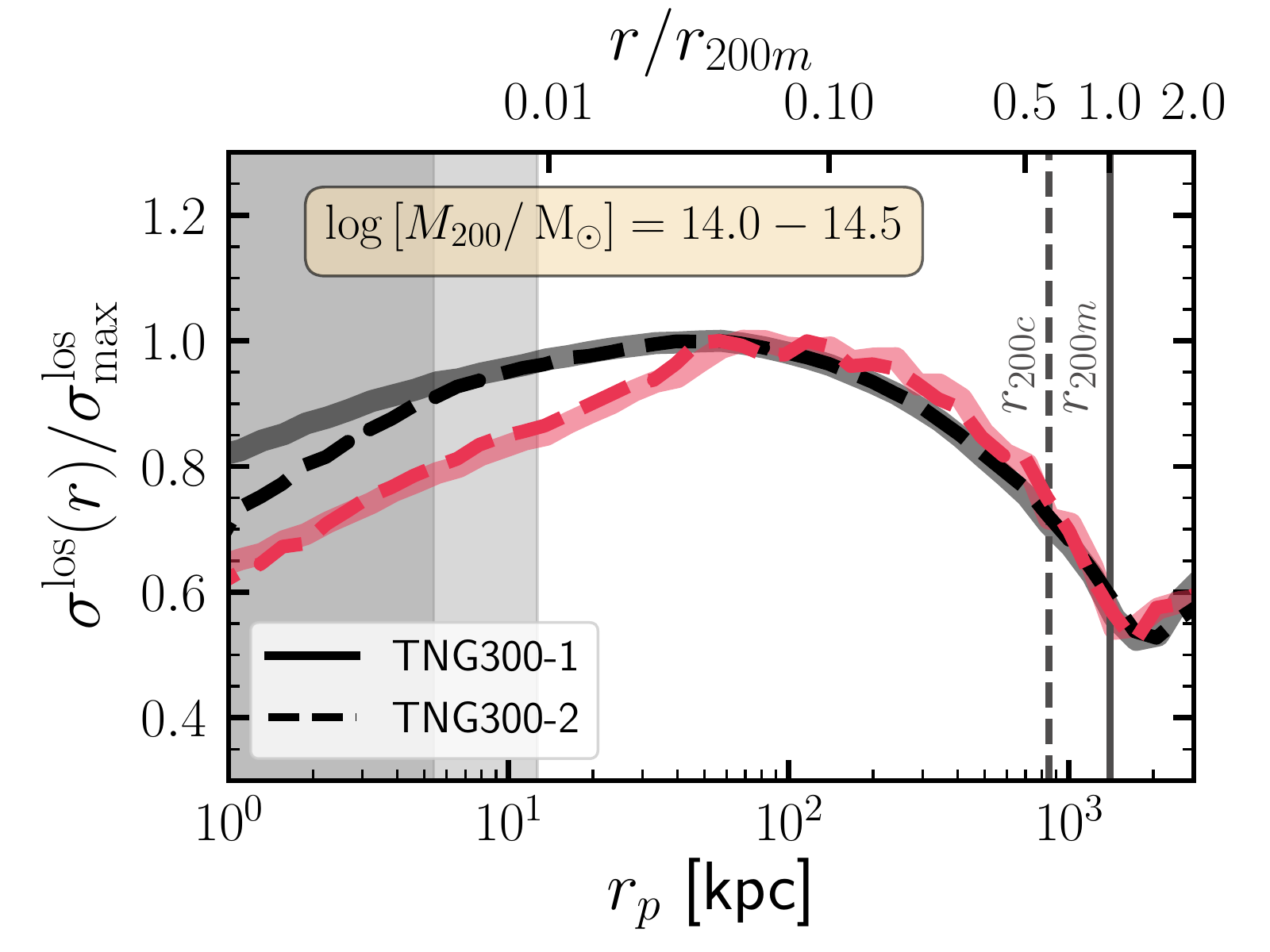}
    \caption{Comparison of the line-of-sight velocity dispersion profiles for dark matter (black) and star particles (red) obtained from the TNG300-1 (solid lines) and TNG300-2 simulations (dashed lines). The shaded gray bands are bounded by the convergence radius appropriate to each simulation set. }
    \label{fig:vdisp_conv}
\end{figure*}

Figure~\ref{fig:vdisp_conv} compares the mean (normalized) velocity dispersion profiles of dark matter and star particles from each of these simulations across two mass bins: $\log\left[M_{200}/{\rm M}_\odot\right]=13.0-13.5$ (left panel) and $\log\left[M_{200}/{\rm M}_\odot\right]=14.0-14.5$ (right panel). The solid lines show results from TNG300-1, while the dashed curves are the corresponding results from TNG300-2. The gray shaded regions mark the convergence radii at each of the resolution levels. The light gray band (which extends to larger radii) corresponds to TNG300-2, and specifies the minimum radius down to which the profiles can be expected to converge given the numerical settings adopted in the simulation.

In general, we find excellent agreement in the dispersion profiles extracted from the two versions of the TNG300 simulation, particularly in the region outside the convergence radius of the TNG300-2 simulation. Not only is the general shape captured equally well (in both dark matter and stars), we also find that the location of the peak and the overall concavity of the profile (which, respectively, define $r_\star$ and $\chi_\star$ in Eq.~\ref{eq:disp_eq}) are well matched. Finally, we also note that the kink in the outer profile is present at both resolution levels, and occurs at roughly the same location ($r_p\approx1.5r_{200m}$). We also find that the level of agreement is, in general, much better in the case of the {\it normalized} dispersion profiles (i.e., where, at a given resolution level, $\sigma^{{\rm los}}(r_p)$ is rescaled by $\sigma^{{\rm los}}_{{\rm max}}$) than when comparing the absolute magnitude of the velocity dispersion profiles.

\section{The radial velocity dispersion profile}
\label{sec:radialdisp}

\begin{figure*}
\centering
\includegraphics[width=0.475\textwidth]{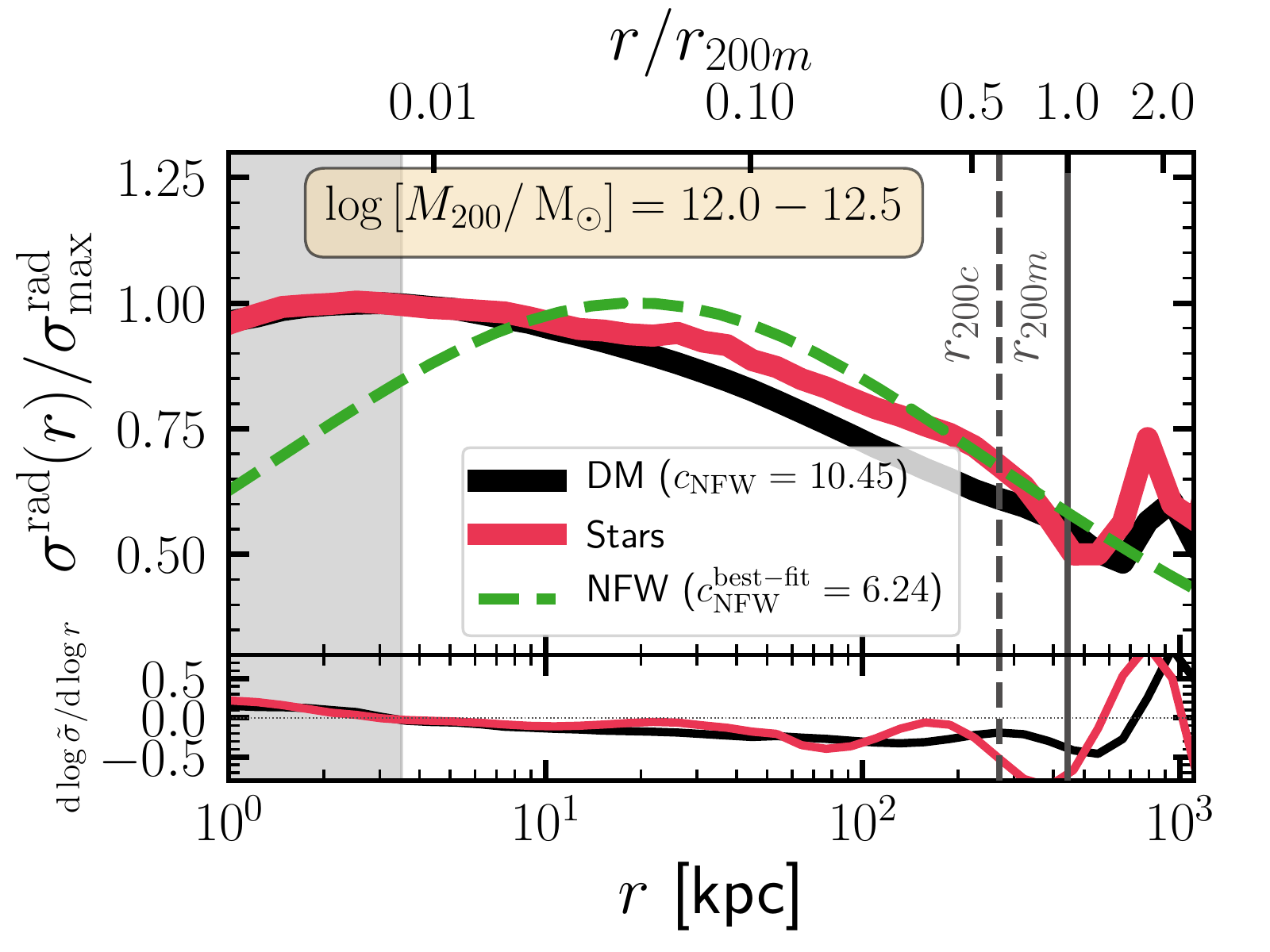}
\includegraphics[width=0.475\textwidth]{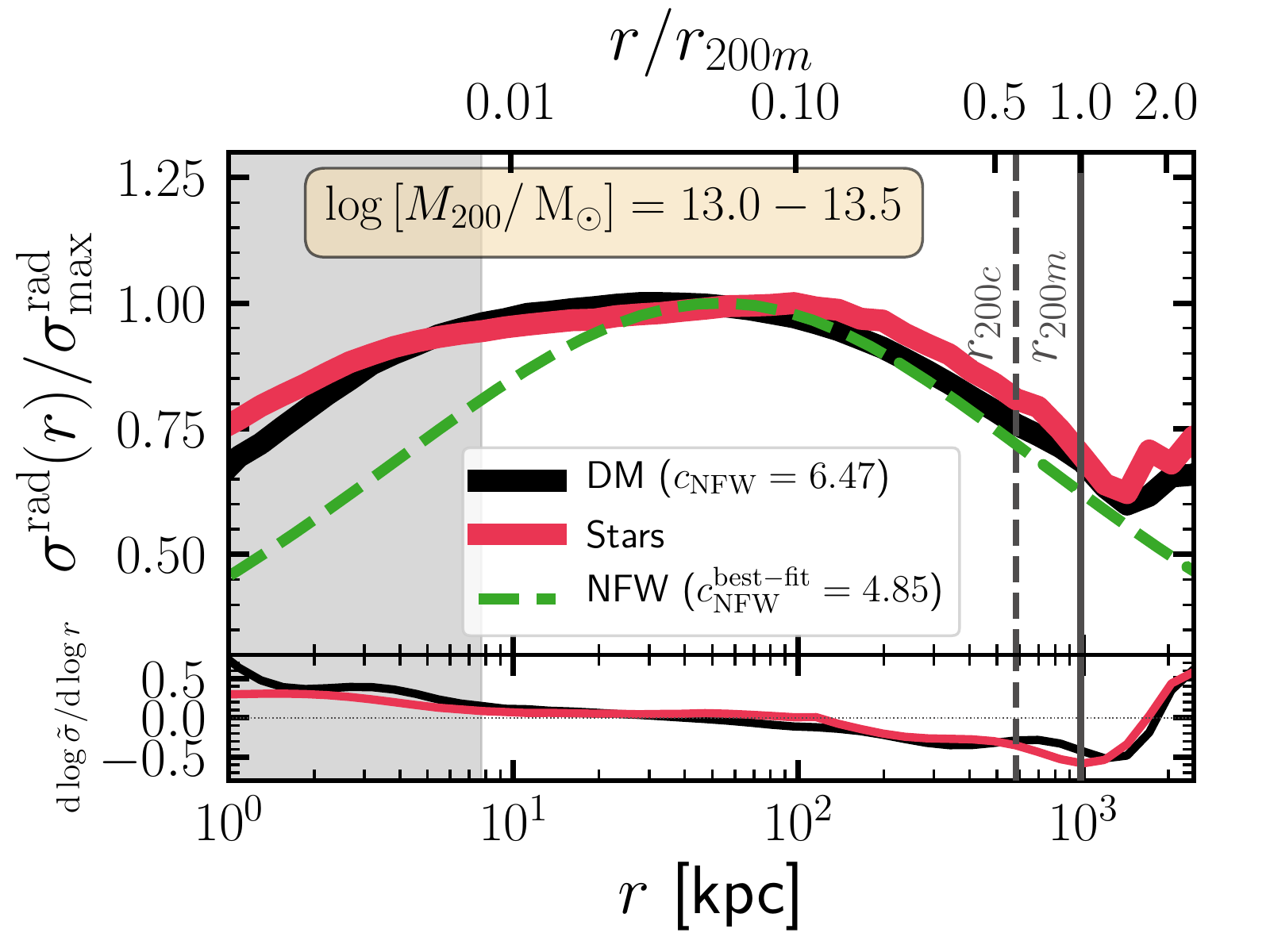}\\ 
\includegraphics[width=0.475\textwidth]{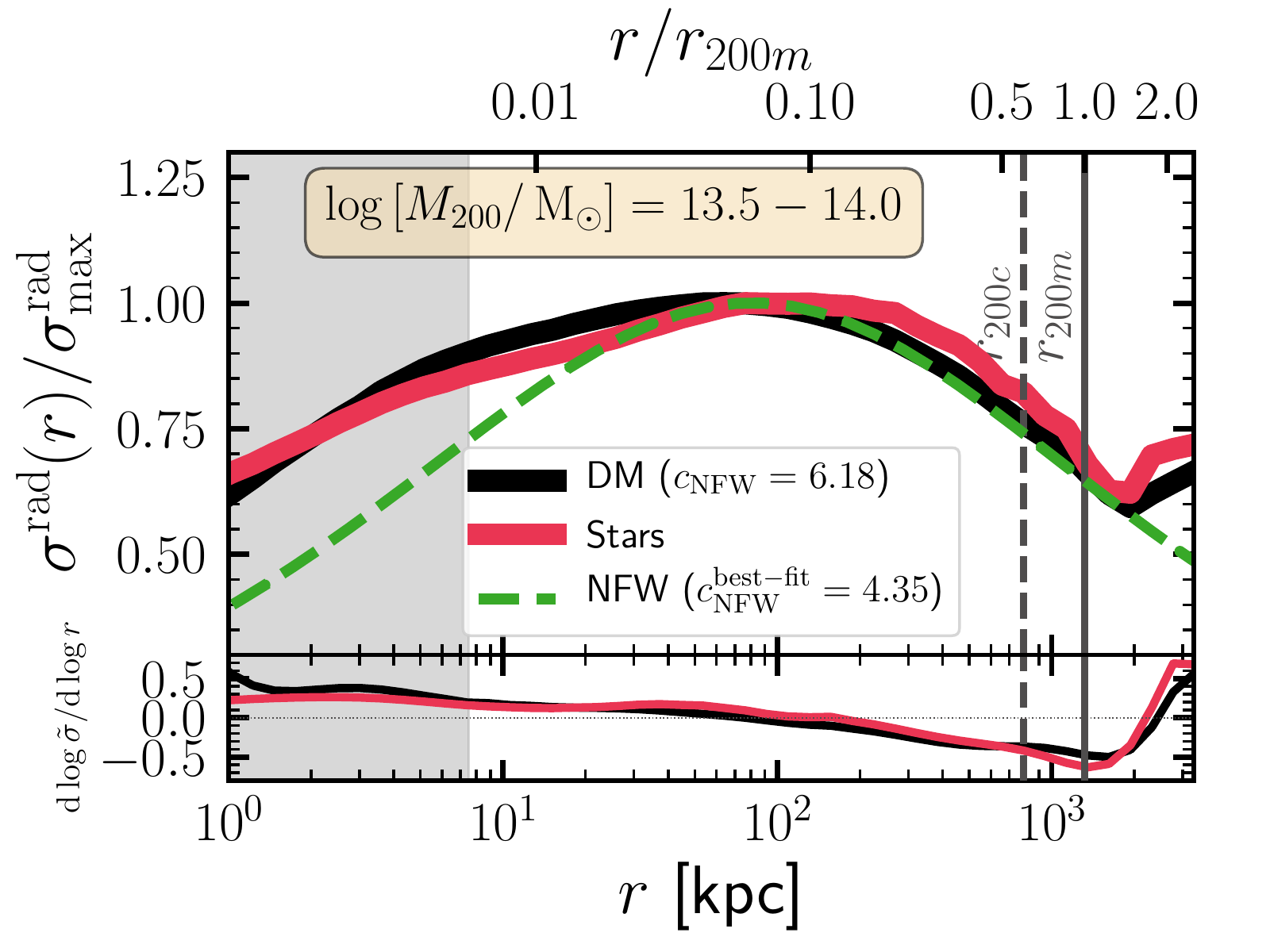}
\includegraphics[width=0.475\textwidth]{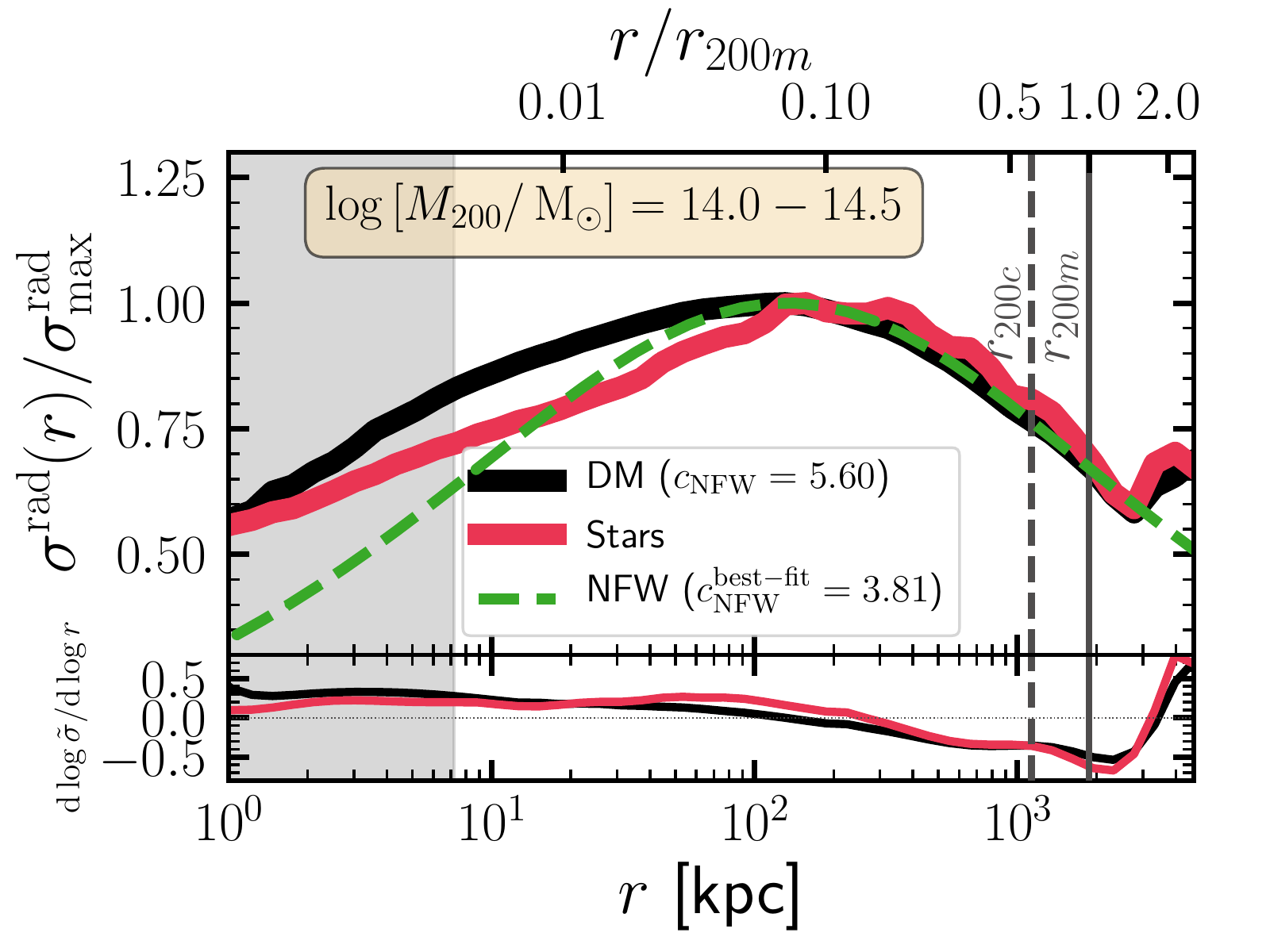}\\
\caption{The stacked {\it radial} velocity dispersion profiles of dark matter (black) and star particles (red) in halos identified from TNG. The colors and linestyles are identical to that in Figure~\ref{fig:los_disp_fit}. The dashed green curve shows the shape of the best-fitting radial velocity dispersion profile as predicted by the NFW profile. While the analytic prediction does a sufficiently good job at capturing the behavior of the TNG profiles on scales $r\gtrsim0.1 r_{200m}$, the agreement worsens towards the centers of halos, where the baryonic content dominates the total gravitational potential and can impact the velocity dispersion of the dark matter relative to a DMO simulation.}
\label{fig:radial_disp}
\end{figure*}

Our main focus in this paper has been on the nature of the velocity dispersion profiles of TNG halos as measured by their line-of-sight motions. In this Appendix, we also present the radial velocity dispersion profiles of these objects; the results are shown in Figure~\ref{fig:radial_disp}. Each panel shows the stacked radial profiles in bins of halo mass, with the colors and linestyles identical to that in Figure~\ref{fig:los_disp_fit}.

We notice immediately that the normalized radial velocity dispersion profiles exhibit a number of similar features to the line-of-sight dispersion profiles examined in Figure~\ref{fig:los_disp_fit}. In particular, the kink in the outer profile is also imprinted in the radial velocities, while the average location of $r_{200m}$ can once again be identified as the radial distance at which the velocity dispersion falls to 60\% of its maximum value. 

The dashed green line in each panel shows the radial velocity dispersion profile for NFW halos of the corresponding mass. Each curve is computed as a solution to the isotropic Jeans equation, giving:
\begin{equation} \label{eq:jeans_radial}
    \sigma_{{\rm rad}}^2(r) = V_{200}^2\frac{c_{{\rm 200}}}{g(c_{{\rm 200}})} \left( \frac{r}{r_s} \right) \left(1+\frac{r}{r_s}\right)^2 \int^\infty_{r/r_s} {\rm d} x \frac{g(x)}{x^3\left(1+x\right)^2},
\end{equation}
where $g(y) \equiv \ln (1+y) - \frac{y}{1+y}$, $r_s$ is the NFW scale radius, and $V_{200}$ is the circular velocity measured at the halo virial radius. The concentration, $c_{{\rm 200}}$, is allowed to vary as a free parameter; the value that provides the best agreement between the theoretical prediction (Eq.~\ref{eq:jeans_radial}) and the dark matter radial velocity dispersion profiles in Figure~\ref{fig:radial_disp} is quoted as $c_{{\rm NFW}}^{{\rm best-fit}}$ in each panel. Note that this value is consistently lower than the value of the (median) concentration inferred from fitting NFW profiles directly to the dark matter density profiles of individual halos in each mass bin (quoted simply as $c_{{\rm NFW}}$ in each panel). In general, we find that $c_{{\rm NFW}}^{{\rm best-fit}} \approx 0.8 c_{200}$, where $c_{200}$ is the concentration of the halo predicted using the \cite{Ludlow2016} model, which was also the definition used in Eq.~\ref{eq:disp_eq_recast}. 

Eq.~(\ref{eq:jeans_radial}), which is generally found to provide an excellent fit to the radial velocity dispersion profiles of halos extracted from dark matter-only simulations \citep[see, e.g.][]{Diemer2013}, does reasonably well in describing the profiles from the full-physics TNG halos in the regime $r\gtrsim0.1r_{200m}$. Internal to this radius, however, there is a significant discrepancy between the measurements from the simulation and the NFW prediction. This is not unexpected, since it has been known for some time that the presence of baryons has a non-negligible impact on the inner velocity structure of dark matter halos. In particular, the velocity dispersion in the inner regions of halos in hydrodynamical simulations are larger than their counterparts in dark matter-only simulations \citep[e.g.][]{Pedrosa2010,Tissera2010}. While the NFW profile predicts a sharp temperature inversion at the centers of halos, the presence of baryons increases the velocity dispersion, broadens the profile below $\sim 0.05 r_{200m}$, and also isotropizes particle orbits in the inner halo \citep[see e.g.][for comparisons made using the Illustris simulation]{Chua2019}.

\bibliography{refs}{} \bibliographystyle{aasjournal}

\begin{thebibliography}{}
\expandafter\ifx\csname natexlab\endcsname\relax\def\natexlab#1{#1}\fi
\providecommand{\url}[1]{\href{#1}{#1}}
\providecommand{\dodoi}[1]{doi:~\href{http://doi.org/#1}{\nolinkurl{#1}}}
\providecommand{\doeprint}[1]{\href{http://ascl.net/#1}{\nolinkurl{http://ascl.net/#1}}}
\providecommand{\doarXiv}[1]{\href{https://arxiv.org/abs/#1}{\nolinkurl{https://arxiv.org/abs/#1}}}

\bibitem[{{Adhikari} {et~al.}(2014){Adhikari}, {Dalal}, \&
  {Chamberlain}}]{Adhikari2014}
{Adhikari}, S., {Dalal}, N., \& {Chamberlain}, R.~T. 2014, \jcap, 2014, 019,
  \dodoi{10.1088/1475-7516/2014/11/019}

\bibitem[{{Adhikari} {et~al.}(2018){Adhikari}, {Sakstein}, {Jain}, {Dalal}, \&
  {Li}}]{Adhikari2018}
{Adhikari}, S., {Sakstein}, J., {Jain}, B., {Dalal}, N., \& {Li}, B. 2018,
  \jcap, 2018, 033, \dodoi{10.1088/1475-7516/2018/11/033}

\bibitem[{{Baxter} {et~al.}(2017){Baxter}, {Chang}, {Jain}, {Adhikari},
  {Dalal}, {Kravtsov}, {More}, {Rozo}, {Rykoff}, \& {Sheth}}]{Baxter2017}
{Baxter}, E., {Chang}, C., {Jain}, B., {et~al.} 2017, \apj, 841, 18,
  \dodoi{10.3847/1538-4357/aa6ff0}

\bibitem[{{Bertschinger}(1985)}]{Bertschinger1985}
{Bertschinger}, E. 1985, \apjs, 58, 39, \dodoi{10.1086/191028}

\bibitem[{{Bianconi} {et~al.}(2020){Bianconi}, {Buscicchio}, {Smith}, {McGee},
  {Haines}, {Finoguenov}, \& {Babul}}]{Bianconi2020}
{Bianconi}, M., {Buscicchio}, R., {Smith}, G.~P., {et~al.} 2020, arXiv
  e-prints, arXiv:2010.05920.
\newblock \doarXiv{2010.05920}

\bibitem[{{Binney} \& {Mamon}(1982)}]{Binney1982}
{Binney}, J., \& {Mamon}, G.~A. 1982, \mnras, 200, 361,
  \dodoi{10.1093/mnras/200.2.361}

\bibitem[{{Brown} {et~al.}(2010){Brown}, {Geller}, {Kenyon}, \&
  {Diaferio}}]{Brown2010}
{Brown}, W.~R., {Geller}, M.~J., {Kenyon}, S.~J., \& {Diaferio}, A. 2010, \aj,
  139, 59, \dodoi{10.1088/0004-6256/139/1/59}

\bibitem[{{Bryan} \& {Norman}(1998)}]{Bryan1998}
{Bryan}, G.~L., \& {Norman}, M.~L. 1998, \apj, 495, 80, \dodoi{10.1086/305262}

\bibitem[{{Bullock} {et~al.}(2001){Bullock}, {Kolatt}, {Sigad}, {Somerville},
  {Kravtsov}, {Klypin}, {Primack}, \& {Dekel}}]{Bullock2001}
{Bullock}, J.~S., {Kolatt}, T.~S., {Sigad}, Y., {et~al.} 2001, \mnras, 321,
  559, \dodoi{10.1046/j.1365-8711.2001.04068.x}

\bibitem[{{Busch} \& {White}(2017)}]{Busch2017}
{Busch}, P., \& {White}, S. D.~M. 2017, \mnras, 470, 4767,
  \dodoi{10.1093/mnras/stx1584}

\bibitem[{{Calderon} \& {Berlind}(2019)}]{Calderon2019}
{Calderon}, V.~F., \& {Berlind}, A.~A. 2019, \mnras, 490, 2367,
  \dodoi{10.1093/mnras/stz2775}

\bibitem[{{Chang} {et~al.}(2018){Chang}, {Baxter}, {Jain}, {S{\'a}nchez},
  {Adhikari}, {Varga}, {Fang}, {Rozo}, {Rykoff}, {Kravtsov}, {Gruen},
  {Hartley}, {Huff}, {Jarvis}, {Kim}, {Prat}, {MacCrann}, {McClintock},
  {Palmese}, {Rapetti}, {Rollins}, {Samuroff}, {Sheldon}, {Troxel}, {Wechsler},
  {Zhang}, {Zuntz}, {Abbott}, {Abdalla}, {Allam}, {Annis}, {Bechtol},
  {Benoit-L{\'e}vy}, {Bernstein}, {Brooks}, {Buckley-Geer}, {Carnero Rosell},
  {Carrasco Kind}, {Carretero}, {D'Andrea}, {da Costa}, {Davis}, {Desai},
  {Diehl}, {Dietrich}, {Drlica-Wagner}, {Eifler}, {Flaugher}, {Fosalba},
  {Frieman}, {Garc{\'\i}a-Bellido}, {Gaztanaga}, {Gerdes}, {Gruendl},
  {Gschwend}, {Gutierrez}, {Honscheid}, {James}, {Jeltema}, {Krause}, {Kuehn},
  {Lahav}, {Lima}, {March}, {Marshall}, {Martini}, {Melchior}, {Menanteau},
  {Miquel}, {Mohr}, {Nord}, {Ogando}, {Plazas}, {Sanchez}, {Scarpine},
  {Schindler}, {Schubnell}, {Sevilla-Noarbe}, {Smith}, {Smith},
  {Soares-Santos}, {Sobreira}, {Suchyta}, {Swanson}, {Tarle}, {Weller}, \& {DES
  Collaboration}}]{Chang2018}
{Chang}, C., {Baxter}, E., {Jain}, B., {et~al.} 2018, \apj, 864, 83,
  \dodoi{10.3847/1538-4357/aad5e7}

\bibitem[{{Chua} {et~al.}(2019){Chua}, {Pillepich}, {Vogelsberger}, \&
  {Hernquist}}]{Chua2019}
{Chua}, K. T.~E., {Pillepich}, A., {Vogelsberger}, M., \& {Hernquist}, L. 2019,
  \mnras, 484, 476, \dodoi{10.1093/mnras/sty3531}

\bibitem[{{Cohen} {et~al.}(2017){Cohen}, {Sesar}, {Bahnolzer}, {He},
  {Kulkarni}, {Prince}, {Bellm}, \& {Laher}}]{Cohen2017}
{Cohen}, J.~G., {Sesar}, B., {Bahnolzer}, S., {et~al.} 2017, \apj, 849, 150,
  \dodoi{10.3847/1538-4357/aa9120}

\bibitem[{{Davis} {et~al.}(1985){Davis}, {Efstathiou}, {Frenk}, \&
  {White}}]{Davis1985}
{Davis}, M., {Efstathiou}, G., {Frenk}, C.~S., \& {White}, S.~D.~M. 1985, \apj,
  292, 371, \dodoi{10.1086/163168}

\bibitem[{{Deason} {et~al.}(2020{\natexlab{a}}){Deason}, {Fattahi}, {Frenk},
  {Grand }, {Oman}, {Garrison-Kimmel}, {Simpson}, \& {Navarro}}]{Deason2020}
{Deason}, A.~J., {Fattahi}, A., {Frenk}, C.~S., {et~al.} 2020{\natexlab{a}},
  \mnras, 496, 3929, \dodoi{10.1093/mnras/staa1711}

\bibitem[{{Deason} {et~al.}(2012){Deason}, {Belokurov}, {Evans}, {Koposov},
  {Cooke}, {Pe{\~n}arrubia}, {Laporte}, {Fellhauer}, {Walker}, \&
  {Olszewski}}]{Deason2012}
{Deason}, A.~J., {Belokurov}, V., {Evans}, N.~W., {et~al.} 2012, \mnras, 425,
  2840, \dodoi{10.1111/j.1365-2966.2012.21639.x}

\bibitem[{{Deason} {et~al.}(2020{\natexlab{b}}){Deason}, {Oman}, {Fattahi},
  {Schaller}, {Jauzac}, {Zhang}, {Montes}, {Bah{\'e}}, {Dalla Vecchia}, {Kay},
  \& {Evans}}]{Deason2020b}
{Deason}, A.~J., {Oman}, K.~A., {Fattahi}, A., {et~al.} 2020{\natexlab{b}},
  arXiv e-prints, arXiv:2010.02937.
\newblock \doarXiv{2010.02937}

\bibitem[{{Diemer}(2017)}]{Diemer2017}
{Diemer}, B. 2017, \apjs, 231, 5, \dodoi{10.3847/1538-4365/aa799c}

\bibitem[{{Diemer} \& {Kravtsov}(2014)}]{Diemer2014}
{Diemer}, B., \& {Kravtsov}, A.~V. 2014, \apj, 789, 1,
  \dodoi{10.1088/0004-637X/789/1/1}

\bibitem[{{Diemer} {et~al.}(2013){Diemer}, {Kravtsov}, \& {More}}]{Diemer2013}
{Diemer}, B., {Kravtsov}, A.~V., \& {More}, S. 2013, \apj, 779, 159,
  \dodoi{10.1088/0004-637X/779/2/159}

\bibitem[{{Diemer} {et~al.}(2017){Diemer}, {Mansfield}, {Kravtsov}, \&
  {More}}]{Diemer2017b}
{Diemer}, B., {Mansfield}, P., {Kravtsov}, A.~V., \& {More}, S. 2017, \apj,
  843, 140, \dodoi{10.3847/1538-4357/aa79ab}

\bibitem[{{Dutton} \& {Macci{\`o}}(2014)}]{Dutton2014}
{Dutton}, A.~A., \& {Macci{\`o}}, A.~V. 2014, \mnras, 441, 3359,
  \dodoi{10.1093/mnras/stu742}

\bibitem[{{Eke} {et~al.}(1996){Eke}, {Cole}, \& {Frenk}}]{Eke1996}
{Eke}, V.~R., {Cole}, S., \& {Frenk}, C.~S. 1996, \mnras, 282, 263,
  \dodoi{10.1093/mnras/282.1.263}

\bibitem[{{Fillmore} \& {Goldreich}(1984)}]{Fillmore1984}
{Fillmore}, J.~A., \& {Goldreich}, P. 1984, \apj, 281, 1,
  \dodoi{10.1086/162070}

\bibitem[{{Fong} \& {Han}(2020)}]{Fong2020}
{Fong}, M., \& {Han}, J. 2020, arXiv e-prints, arXiv:2008.03477.
\newblock \doarXiv{2008.03477}

\bibitem[{{Gunn} \& {Gott}(1972)}]{Gunn1972}
{Gunn}, J.~E., \& {Gott}, J.~Richard, I. 1972, \apj, 176, 1,
  \dodoi{10.1086/151605}

\bibitem[{Harris {et~al.}(2020)Harris, Millman, van~der Walt, Gommers,
  Virtanen, Cournapeau, Wieser, Taylor, Berg, Smith, Kern, Picus, Hoyer, van
  Kerkwijk, Brett, Haldane, Fernández~del Río, Wiebe, Peterson,
  Gérard-Marchant, Sheppard, Reddy, Weckesser, Abbasi, Gohlke, \&
  Oliphant}]{Numpy2020}
Harris, C.~R., Millman, K.~J., van~der Walt, S.~J., {et~al.} 2020, Nature, 585,
  357–362, \dodoi{10.1038/s41586-020-2649-2}

\bibitem[{Hunter(2007)}]{Hunter2007}
Hunter, J.~D. 2007, Computing in Science \& Engineering, 9, 90,
  \dodoi{10.1109/MCSE.2007.55}

\bibitem[{{Ivezi{\'c}} {et~al.}(2019){Ivezi{\'c}}, {Kahn}, {Tyson}, {Abel},
  {Acosta}, {Allsman}, {Alonso}, {AlSayyad}, {Anderson}, {Andrew}, {Angel},
  {Angeli}, {Ansari}, {Antilogus}, {Araujo}, {Armstrong}, {Arndt}, {Astier},
  {Aubourg}, {Auza}, {Axelrod}, {Bard}, {Barr}, {Barrau}, {Bartlett}, {Bauer},
  {Bauman}, {Baumont}, {Bechtol}, {Bechtol}, {Becker}, {Becla}, {Beldica},
  {Bellavia}, {Bianco}, {Biswas}, {Blanc}, {Blazek}, {Bland ford}, {Bloom},
  {Bogart}, {Bond}, {Booth}, {Borgland}, {Borne}, {Bosch}, {Boutigny},
  {Brackett}, {Bradshaw}, {Brand t}, {Brown}, {Bullock}, {Burchat}, {Burke},
  {Cagnoli}, {Calabrese}, {Callahan}, {Callen}, {Carlin}, {Carlson}, {Chand
  rasekharan}, {Charles-Emerson}, {Chesley}, {Cheu}, {Chiang}, {Chiang},
  {Chirino}, {Chow}, {Ciardi}, {Claver}, {Cohen-Tanugi}, {Cockrum}, {Coles},
  {Connolly}, {Cook}, {Cooray}, {Covey}, {Cribbs}, {Cui}, {Cutri}, {Daly},
  {Daniel}, {Daruich}, {Daubard}, {Daues}, {Dawson}, {Delgado}, {Dellapenna},
  {de Peyster}, {de Val-Borro}, {Digel}, {Doherty}, {Dubois},
  {Dubois-Felsmann}, {Durech}, {Economou}, {Eifler}, {Eracleous}, {Emmons},
  {Fausti Neto}, {Ferguson}, {Figueroa}, {Fisher-Levine}, {Focke}, {Foss},
  {Frank}, {Freemon}, {Gangler}, {Gawiser}, {Geary}, {Gee}, {Geha}, {Gessner},
  {Gibson}, {Gilmore}, {Glanzman}, {Glick}, {Goldina}, {Goldstein}, {Goodenow},
  {Graham}, {Gressler}, {Gris}, {Guy}, {Guyonnet}, {Haller}, {Harris},
  {Hascall}, {Haupt}, {Hernand ez}, {Herrmann}, {Hileman}, {Hoblitt},
  {Hodgson}, {Hogan}, {Howard}, {Huang}, {Huffer}, {Ingraham}, {Innes},
  {Jacoby}, {Jain}, {Jammes}, {Jee}, {Jenness}, {Jernigan}, {Jevremovi{\'c}},
  {Johns}, {Johnson}, {Johnson}, {Jones}, {Juramy-Gilles}, {Juri{\'c}},
  {Kalirai}, {Kallivayalil}, {Kalmbach}, {Kantor}, {Karst}, {Kasliwal},
  {Kelly}, {Kessler}, {Kinnison}, {Kirkby}, {Knox}, {Kotov}, {Krabbendam},
  {Krughoff}, {Kub{\'a}nek}, {Kuczewski}, {Kulkarni}, {Ku}, {Kurita}, {Lage},
  {Lambert}, {Lange}, {Langton}, {Le Guillou}, {Levine}, {Liang}, {Lim},
  {Lintott}, {Long}, {Lopez}, {Lotz}, {Lupton}, {Lust}, {MacArthur}, {Mahabal},
  {Mand elbaum}, {Markiewicz}, {Marsh}, {Marshall}, {Marshall}, {May},
  {McKercher}, {McQueen}, {Meyers}, {Migliore}, {Miller}, {Mills}, {Miraval},
  {Moeyens}, {Moolekamp}, {Monet}, {Moniez}, {Monkewitz}, {Montgomery},
  {Morrison}, {Mueller}, {Muller}, {Mu{\~n}oz Arancibia}, {Neill}, {Newbry},
  {Nief}, {Nomerotski}, {Nordby}, {O'Connor}, {Oliver}, {Olivier}, {Olsen},
  {O'Mullane}, {Ortiz}, {Osier}, {Owen}, {Pain}, {Palecek}, {Parejko},
  {Parsons}, {Pease}, {Peterson}, {Peterson}, {Petravick}, {Libby Petrick},
  {Petry}, {Pierfederici}, {Pietrowicz}, {Pike}, {Pinto}, {Plante}, {Plate},
  {Plutchak}, {Price}, {Prouza}, {Radeka}, {Rajagopal}, {Rasmussen},
  {Regnault}, {Reil}, {Reiss}, {Reuter}, {Ridgway}, {Riot}, {Ritz}, {Robinson},
  {Roby}, {Roodman}, {Rosing}, {Roucelle}, {Rumore}, {Russo}, {Saha},
  {Sassolas}, {Schalk}, {Schellart}, {Schindler}, {Schmidt}, {Schneider},
  {Schneider}, {Schoening}, {Schumacher}, {Schwamb}, {Sebag}, {Selvy},
  {Sembroski}, {Seppala}, {Serio}, {Serrano}, {Shaw}, {Shipsey}, {Sick},
  {Silvestri}, {Slater}, {Smith}, {Smith}, {Sobhani}, {Soldahl},
  {Storrie-Lombardi}, {Stover}, {Strauss}, {Street}, {Stubbs}, {Sullivan},
  {Sweeney}, {Swinbank}, {Szalay}, {Takacs}, {Tether}, {Thaler}, {Thayer},
  {Thomas}, {Thornton}, {Thukral}, {Tice}, {Trilling}, {Turri}, {Van Berg},
  {Vanden Berk}, {Vetter}, {Virieux}, {Vucina}, {Wahl}, {Walkowicz}, {Walsh},
  {Walter}, {Wang}, {Wang}, {Warner}, {Wiecha}, {Willman}, {Winters},
  {Wittman}, {Wolff}, {Wood-Vasey}, {Wu}, {Xin}, {Yoachim}, \&
  {Zhan}}]{Ivezic2019}
{Ivezi{\'c}}, {\v{Z}}., {Kahn}, S.~M., {Tyson}, J.~A., {et~al.} 2019, \apj,
  873, 111, \dodoi{10.3847/1538-4357/ab042c}

\bibitem[{{Kaviraj}(2020)}]{Kaviraj2020}
{Kaviraj}, S. 2020, arXiv e-prints, arXiv:2001.01728.
\newblock \doarXiv{2001.01728}

\bibitem[{{Kravtsov} {et~al.}(2004){Kravtsov}, {Berlind}, {Wechsler}, {Klypin},
  {Gottl{\"o}ber}, {Allgood}, \& {Primack}}]{Kravtsov2004}
{Kravtsov}, A.~V., {Berlind}, A.~A., {Wechsler}, R.~H., {et~al.} 2004, \apj,
  609, 35, \dodoi{10.1086/420959}

\bibitem[{{Laureijs} {et~al.}(2011){Laureijs}, {Amiaux}, {Arduini},
  {Augu{\`e}res}, {Brinchmann}, {Cole}, {Cropper}, {Dabin}, {Duvet}, {Ealet},
  {Garilli}, {Gondoin}, {Guzzo}, {Hoar}, {Hoekstra}, {Holmes}, {Kitching},
  {Maciaszek}, {Mellier}, {Pasian}, {Percival}, {Rhodes}, {Saavedra Criado},
  {Sauvage}, {Scaramella}, {Valenziano}, {Warren}, {Bender}, {Castander},
  {Cimatti}, {Le F{\`e}vre}, {Kurki-Suonio}, {Levi}, {Lilje}, {Meylan},
  {Nichol}, {Pedersen}, {Popa}, {Rebolo Lopez}, {Rix}, {Rottgering},
  {Zeilinger}, {Grupp}, {Hudelot}, {Massey}, {Meneghetti}, {Miller}, {Paltani},
  {Paulin-Henriksson}, {Pires}, {Saxton}, {Schrabback}, {Seidel}, {Walsh},
  {Aghanim}, {Amendola}, {Bartlett}, {Baccigalupi}, {Beaulieu}, {Benabed},
  {Cuby}, {Elbaz}, {Fosalba}, {Gavazzi}, {Helmi}, {Hook}, {Irwin}, {Kneib},
  {Kunz}, {Mannucci}, {Moscardini}, {Tao}, {Teyssier}, {Weller}, {Zamorani},
  {Zapatero Osorio}, {Boulade}, {Foumond}, {Di Giorgio}, {Guttridge}, {James},
  {Kemp}, {Martignac}, {Spencer}, {Walton}, {Bl{\"u}mchen}, {Bonoli},
  {Bortoletto}, {Cerna}, {Corcione}, {Fabron}, {Jahnke}, {Ligori}, {Madrid},
  {Martin}, {Morgante}, {Pamplona}, {Prieto}, {Riva}, {Toledo}, {Trifoglio},
  {Zerbi}, {Abdalla}, {Douspis}, {Grenet}, {Borgani}, {Bouwens}, {Courbin},
  {Delouis}, {Dubath}, {Fontana}, {Frailis}, {Grazian}, {Koppenh{\"o}fer},
  {Mansutti}, {Melchior}, {Mignoli}, {Mohr}, {Neissner}, {Noddle}, {Poncet},
  {Scodeggio}, {Serrano}, {Shane}, {Starck}, {Surace}, {Taylor},
  {Verdoes-Kleijn}, {Vuerli}, {Williams}, {Zacchei}, {Altieri}, {Escudero
  Sanz}, {Kohley}, {Oosterbroek}, {Astier}, {Bacon}, {Bardelli}, {Baugh},
  {Bellagamba}, {Benoist}, {Bianchi}, {Biviano}, {Branchini}, {Carbone},
  {Cardone}, {Clements}, {Colombi}, {Conselice}, {Cresci}, {Deacon}, {Dunlop},
  {Fedeli}, {Fontanot}, {Franzetti}, {Giocoli}, {Garcia-Bellido}, {Gow},
  {Heavens}, {Hewett}, {Heymans}, {Holland}, {Huang}, {Ilbert}, {Joachimi},
  {Jennins}, {Kerins}, {Kiessling}, {Kirk}, {Kotak}, {Krause}, {Lahav}, {van
  Leeuwen}, {Lesgourgues}, {Lombardi}, {Magliocchetti}, {Maguire}, {Majerotto},
  {Maoli}, {Marulli}, {Maurogordato}, {McCracken}, {McLure}, {Melchiorri},
  {Merson}, {Moresco}, {Nonino}, {Norberg}, {Peacock}, {Pello}, {Penny},
  {Pettorino}, {Di Porto}, {Pozzetti}, {Quercellini}, {Radovich}, {Rassat},
  {Roche}, {Ronayette}, {Rossetti}, {Sartoris}, {Schneider}, {Semboloni},
  {Serjeant}, {Simpson}, {Skordis}, {Smadja}, {Smartt}, {Spano}, {Spiro},
  {Sullivan}, {Tilquin}, {Trotta}, {Verde}, {Wang}, {Williger}, {Zhao},
  {Zoubian}, \& {Zucca}}]{Laureijs2011}
{Laureijs}, R., {Amiaux}, J., {Arduini}, S., {et~al.} 2011, arXiv e-prints,
  arXiv:1110.3193.
\newblock \doarXiv{1110.3193}

\bibitem[{{{\L}okas} \& {Mamon}(2001)}]{Lokas2001}
{{\L}okas}, E.~L., \& {Mamon}, G.~A. 2001, \mnras, 321, 155,
  \dodoi{10.1046/j.1365-8711.2001.04007.x}

\bibitem[{{Lovell} {et~al.}(2018){Lovell}, {Pillepich}, {Genel}, {Nelson},
  {Springel}, {Pakmor}, {Marinacci}, {Weinberger}, {Torrey}, {Vogelsberger},
  {Alabi}, \& {Hernquist}}]{Lovell2018}
{Lovell}, M.~R., {Pillepich}, A., {Genel}, S., {et~al.} 2018, \mnras, 481,
  1950, \dodoi{10.1093/mnras/sty2339}

\bibitem[{{Ludlow} {et~al.}(2016){Ludlow}, {Bose}, {Angulo}, {Wang},
  {Hellwing}, {Navarro}, {Cole}, \& {Frenk}}]{Ludlow2016}
{Ludlow}, A.~D., {Bose}, S., {Angulo}, R.~E., {et~al.} 2016, \mnras, 460, 1214,
  \dodoi{10.1093/mnras/stw1046}

\bibitem[{{Ludlow} {et~al.}(2013){Ludlow}, {Navarro}, {Boylan-Kolchin}, {Bett},
  {Angulo}, {Li}, {White}, {Frenk}, \& {Springel}}]{Ludlow2013}
{Ludlow}, A.~D., {Navarro}, J.~F., {Boylan-Kolchin}, M., {et~al.} 2013, \mnras,
  432, 1103, \dodoi{10.1093/mnras/stt526}

\bibitem[{{Mansfield} {et~al.}(2017){Mansfield}, {Kravtsov}, \&
  {Diemer}}]{Mansfield2017}
{Mansfield}, P., {Kravtsov}, A.~V., \& {Diemer}, B. 2017, \apj, 841, 34,
  \dodoi{10.3847/1538-4357/aa7047}

\bibitem[{{Marinacci} {et~al.}(2018){Marinacci}, {Vogelsberger}, {Pakmor},
  {Torrey}, {Springel}, {Hernquist}, {Nelson}, {Weinberger}, {Pillepich},
  {Naiman}, \& {Genel}}]{Marinacci2018}
{Marinacci}, F., {Vogelsberger}, M., {Pakmor}, R., {et~al.} 2018, \mnras, 480,
  5113, \dodoi{10.1093/mnras/sty2206}

\bibitem[{{Mogotsi} \& {Romeo}(2019)}]{Mogotsi2019}
{Mogotsi}, K.~M., \& {Romeo}, A.~B. 2019, \mnras, 489, 3797,
  \dodoi{10.1093/mnras/stz2370}

\bibitem[{{More} {et~al.}(2015){More}, {Diemer}, \& {Kravtsov}}]{More2015}
{More}, S., {Diemer}, B., \& {Kravtsov}, A.~V. 2015, \apj, 810, 36,
  \dodoi{10.1088/0004-637X/810/1/36}

\bibitem[{{More} {et~al.}(2009){More}, {van den Bosch}, \&
  {Cacciato}}]{More2009}
{More}, S., {van den Bosch}, F.~C., \& {Cacciato}, M. 2009, \mnras, 392, 917,
  \dodoi{10.1111/j.1365-2966.2008.14114.x}

\bibitem[{{More} {et~al.}(2016){More}, {Miyatake}, {Takada}, {Diemer},
  {Kravtsov}, {Dalal}, {More}, {Murata}, {Mandelbaum}, {Rozo}, {Rykoff},
  {Oguri}, \& {Spergel}}]{More2016}
{More}, S., {Miyatake}, H., {Takada}, M., {et~al.} 2016, \apj, 825, 39,
  \dodoi{10.3847/0004-637X/825/1/39}

\bibitem[{{Murata} {et~al.}(2020){Murata}, {Sunayama}, {Oguri}, {More},
  {Nishizawa}, {Nishimichi}, \& {Osato}}]{Murata2020}
{Murata}, R., {Sunayama}, T., {Oguri}, M., {et~al.} 2020, \pasj, 72, 64,
  \dodoi{10.1093/pasj/psaa041}

\bibitem[{{Naiman} {et~al.}(2018){Naiman}, {Pillepich}, {Springel},
  {Ramirez-Ruiz}, {Torrey}, {Vogelsberger}, {Pakmor}, {Nelson}, {Marinacci},
  {Hernquist}, {Weinberger}, \& {Genel}}]{Naiman2018}
{Naiman}, J.~P., {Pillepich}, A., {Springel}, V., {et~al.} 2018, \mnras, 477,
  1206, \dodoi{10.1093/mnras/sty618}

\bibitem[{{Navarro} {et~al.}(1996){Navarro}, {Frenk}, \& {White}}]{Navarro1996}
{Navarro}, J.~F., {Frenk}, C.~S., \& {White}, S. D.~M. 1996, \apj, 462, 563,
  \dodoi{10.1086/177173}

\bibitem[{{Navarro} {et~al.}(1997){Navarro}, {Frenk}, \& {White}}]{Navarro1997}
---. 1997, \apj, 490, 493, \dodoi{10.1086/304888}

\bibitem[{{Nelson} {et~al.}(2018){Nelson}, {Pillepich}, {Springel},
  {Weinberger}, {Hernquist}, {Pakmor}, {Genel}, {Torrey}, {Vogelsberger},
  {Kauffmann}, {Marinacci}, \& {Naiman}}]{Nelson2018a}
{Nelson}, D., {Pillepich}, A., {Springel}, V., {et~al.} 2018, \mnras, 475, 624,
  \dodoi{10.1093/mnras/stx3040}

\bibitem[{Nelson {et~al.}(2019)Nelson, Springel, Pillepich, Rodriguez-Gomez,
  Torrey, Genel, Vogelsberger, Pakmor, Marinacci, Weinberger, Kelley, Lovell,
  Diemer, \& Hernquist}]{Nelson2018b}
Nelson, D., Springel, V., Pillepich, A., {et~al.} 2019, Computational
  Astrophysics and Cosmology, 6, 2, \dodoi{10.1186/s40668-019-0028-x}

\bibitem[{{Nishizawa} {et~al.}(2018){Nishizawa}, {Oguri}, {Oogi}, {More},
  {Nishimichi}, {Nagashima}, {Lin}, {Mandelbaum}, {Takada}, {Bahcall},
  {Coupon}, {Huang}, {Jian}, {Komiyama}, {Leauthaud}, {Lin}, {Miyatake},
  {Miyazaki}, \& {Tanaka}}]{Nishizawa2018}
{Nishizawa}, A.~J., {Oguri}, M., {Oogi}, T., {et~al.} 2018, \pasj, 70, S24,
  \dodoi{10.1093/pasj/psx106}

\bibitem[{{Patej} \& {Loeb}(2016)}]{Patej2016}
{Patej}, A., \& {Loeb}, A. 2016, \apj, 824, 69,
  \dodoi{10.3847/0004-637X/824/2/69}

\bibitem[{{Pedrosa} {et~al.}(2010){Pedrosa}, {Tissera}, \&
  {Scannapieco}}]{Pedrosa2010}
{Pedrosa}, S., {Tissera}, P.~B., \& {Scannapieco}, C. 2010, \mnras, 402, 776,
  \dodoi{10.1111/j.1365-2966.2009.15951.x}

\bibitem[{{Pillepich} {et~al.}(2018{\natexlab{a}}){Pillepich}, {Springel},
  {Nelson}, {Genel}, {Naiman}, {Pakmor}, {Hernquist}, {Torrey}, {Vogelsberger},
  {Weinberger}, \& {Marinacci}}]{Pillepich2018}
{Pillepich}, A., {Springel}, V., {Nelson}, D., {et~al.} 2018{\natexlab{a}},
  \mnras, 473, 4077, \dodoi{10.1093/mnras/stx2656}

\bibitem[{{Pillepich} {et~al.}(2018{\natexlab{b}}){Pillepich}, {Nelson},
  {Hernquist}, {Springel}, {Pakmor}, {Torrey}, {Weinberger}, {Genel}, {Naiman},
  {Marinacci}, \& {Vogelsberger}}]{Pillepich2018b}
{Pillepich}, A., {Nelson}, D., {Hernquist}, L., {et~al.} 2018{\natexlab{b}},
  \mnras, 475, 648, \dodoi{10.1093/mnras/stx3112}

\bibitem[{{Planck Collaboration} {et~al.}(2016){Planck Collaboration}, {Ade},
  {Aghanim}, {Arnaud}, {Ashdown}, {Aumont}, {Baccigalupi}, {Banday},
  {Barreiro}, {Bartlett}, \& et~al.}]{Planck2016}
{Planck Collaboration}, {Ade}, P.~A.~R., {Aghanim}, N., {et~al.} 2016, \aap,
  594, A13, \dodoi{10.1051/0004-6361/201525830}

\bibitem[{{Power} {et~al.}(2003){Power}, {Navarro}, {Jenkins}, {Frenk},
  {White}, {Springel}, {Stadel}, \& {Quinn}}]{Power2003}
{Power}, C., {Navarro}, J.~F., {Jenkins}, A., {et~al.} 2003, \mnras, 338, 14,
  \dodoi{10.1046/j.1365-8711.2003.05925.x}

\bibitem[{{Rubin} \& {Loeb}(2013)}]{Rubin2013}
{Rubin}, D., \& {Loeb}, A. 2013, \jcap, 2013, 019,
  \dodoi{10.1088/1475-7516/2013/12/019}

\bibitem[{{Shin} {et~al.}(2019){Shin}, {Adhikari}, {Baxter}, {Chang}, {Jain},
  {Battaglia}, {Bleem}, {Bocquet}, {DeRose}, {Gruen}, {Hilton}, {Kravtsov},
  {McClintock}, {Rozo}, {Rykoff}, {Varga}, {Wechsler}, {Wu}, {Zhang}, {Aiola},
  {Allam}, {Bechtol}, {Benson}, {Bertin}, {Bond}, {Brodwin}, {Brooks},
  {Buckley-Geer}, {Burke}, {Carlstrom}, {Carnero Rosell}, {Carrasco Kind},
  {Carretero}, {Castander}, {Choi}, {Cunha}, {Crawford}, {da Costa}, {De
  Vicente}, {Desai}, {Devlin}, {Dietrich}, {Doel}, {Dunkley}, {Eifler},
  {Evrard}, {Flaugher}, {Fosalba}, {Gallardo}, {Garc{\'\i}a-Bellido},
  {Gaztanaga}, {Gerdes}, {Gralla}, {Gruendl}, {Gschwend}, {Gupta}, {Gutierrez},
  {Hartley}, {Hill}, {Ho}, {Hollowood}, {Honscheid}, {Hoyle}, {Huffenberger},
  {Hughes}, {James}, {Jeltema}, {Kim}, {Krause}, {Kuehn}, {Lahav}, {Lima},
  {Madhavacheril}, {Maia}, {Marshall}, {Maurin}, {McMahon}, {Menanteau},
  {Miller}, {Miquel}, {Mohr}, {Naess}, {Nati}, {Newburgh}, {Niemack}, {Ogando},
  {Page}, {Partridge}, {Patil}, {Plazas}, {Rapetti}, {Reichardt}, {Romer},
  {Sanchez}, {Scarpine}, {Schindler}, {Serrano}, {Smith}, {Smith},
  {Soares-Santos}, {Sobreira}, {Staggs}, {Stark}, {Stein}, {Suchyta},
  {Swanson}, {Tarle}, {Thomas}, {van Engelen}, {Wollack}, \& {Xu}}]{Shin2019}
{Shin}, T., {Adhikari}, S., {Baxter}, E.~J., {et~al.} 2019, \mnras, 487, 2900,
  \dodoi{10.1093/mnras/stz1434}

\bibitem[{{Spergel} {et~al.}(2015){Spergel}, {Gehrels}, {Baltay}, {Bennett},
  {Breckinridge}, {Donahue}, {Dressler}, {Gaudi}, {Greene}, {Guyon}, {Hirata},
  {Kalirai}, {Kasdin}, {Macintosh}, {Moos}, {Perlmutter}, {Postman},
  {Rauscher}, {Rhodes}, {Wang}, {Weinberg}, {Benford}, {Hudson}, {Jeong},
  {Mellier}, {Traub}, {Yamada}, {Capak}, {Colbert}, {Masters}, {Penny},
  {Savransky}, {Stern}, {Zimmerman}, {Barry}, {Bartusek}, {Carpenter}, {Cheng},
  {Content}, {Dekens}, {Demers}, {Grady}, {Jackson}, {Kuan}, {Kruk}, {Melton},
  {Nemati}, {Parvin}, {Poberezhskiy}, {Peddie}, {Ruffa}, {Wallace}, {Whipple},
  {Wollack}, \& {Zhao}}]{Spergel2015}
{Spergel}, D., {Gehrels}, N., {Baltay}, C., {et~al.} 2015, arXiv e-prints,
  arXiv:1503.03757.
\newblock \doarXiv{1503.03757}

\bibitem[{{Springel}(2010)}]{Springel2010}
{Springel}, V. 2010, \mnras, 401, 791, \dodoi{10.1111/j.1365-2966.2009.15715.x}

\bibitem[{{Springel} {et~al.}(2001){Springel}, {White}, {Tormen}, \&
  {Kauffmann}}]{Springel2001}
{Springel}, V., {White}, S.~D.~M., {Tormen}, G., \& {Kauffmann}, G. 2001,
  \mnras, 328, 726, \dodoi{10.1046/j.1365-8711.2001.04912.x}

\bibitem[{{Springel} {et~al.}(2018){Springel}, {Pakmor}, {Pillepich},
  {Weinberger}, {Nelson}, {Hernquist}, {Vogelsberger}, {Genel}, {Torrey},
  {Marinacci}, \& {Naiman}}]{Springel2018}
{Springel}, V., {Pakmor}, R., {Pillepich}, A., {et~al.} 2018, \mnras, 475, 676,
  \dodoi{10.1093/mnras/stx3304}

\bibitem[{{Tasitsiomi} {et~al.}(2004){Tasitsiomi}, {Kravtsov}, {Wechsler}, \&
  {Primack}}]{Tasitsiomi2004}
{Tasitsiomi}, A., {Kravtsov}, A.~V., {Wechsler}, R.~H., \& {Primack}, J.~R.
  2004, \apj, 614, 533, \dodoi{10.1086/423784}

\bibitem[{{Tempel} \& {Tenjes}(2006)}]{Tempel2006}
{Tempel}, E., \& {Tenjes}, P. 2006, \mnras, 371, 1269,
  \dodoi{10.1111/j.1365-2966.2006.10741.x}

\bibitem[{{Tissera} {et~al.}(2010){Tissera}, {White}, {Pedrosa}, \&
  {Scannapieco}}]{Tissera2010}
{Tissera}, P.~B., {White}, S. D.~M., {Pedrosa}, S., \& {Scannapieco}, C. 2010,
  \mnras, 406, 922, \dodoi{10.1111/j.1365-2966.2010.16777.x}

\bibitem[{{Tomooka} {et~al.}(2020){Tomooka}, {Rozo}, {Wagoner}, {Aung},
  {Nagai}, \& {Safonova}}]{Tomooka2020}
{Tomooka}, P., {Rozo}, E., {Wagoner}, E.~L., {et~al.} 2020, arXiv e-prints,
  arXiv:2003.11555.
\newblock \doarXiv{2003.11555}

\bibitem[{{Vale} \& {Ostriker}(2004)}]{Vale2004}
{Vale}, A., \& {Ostriker}, J.~P. 2004, \mnras, 353, 189,
  \dodoi{10.1111/j.1365-2966.2004.08059.x}

\bibitem[{{Veale} {et~al.}(2018){Veale}, {Ma}, {Greene}, {Thomas}, {Blakeslee},
  {Walsh}, \& {Ito}}]{Veale2018}
{Veale}, M., {Ma}, C.-P., {Greene}, J.~E., {et~al.} 2018, \mnras, 473, 5446,
  \dodoi{10.1093/mnras/stx2717}

\bibitem[{Virtanen {et~al.}(2020)Virtanen, Gommers, Oliphant, Haberland, Reddy,
  Cournapeau, Burovski, Peterson, Weckesser, Bright, {van der Walt}, Brett,
  Wilson, Millman, Mayorov, Nelson, Jones, Kern, Larson, Carey, Polat, Feng,
  Moore, {VanderPlas}, Laxalde, Perktold, Cimrman, Henriksen, Quintero, Harris,
  Archibald, Ribeiro, Pedregosa, {van Mulbregt}, \& {SciPy 1.0
  Contributors}}]{Scipy2020}
Virtanen, P., Gommers, R., Oliphant, T.~E., {et~al.} 2020, Nature Methods, 17,
  261, \dodoi{10.1038/s41592-019-0686-2}

\bibitem[{{Vogelsberger} {et~al.}(2013){Vogelsberger}, {Genel}, {Sijacki},
  {Torrey}, {Springel}, \& {Hernquist}}]{Vogelsberger2013}
{Vogelsberger}, M., {Genel}, S., {Sijacki}, D., {et~al.} 2013, \mnras, 436,
  3031, \dodoi{10.1093/mnras/stt1789}

\bibitem[{{Vogelsberger} {et~al.}(2014){Vogelsberger}, {Genel}, {Springel},
  {Torrey}, {Sijacki}, {Xu}, {Snyder}, {Nelson}, \&
  {Hernquist}}]{Vogelsberger2014a}
{Vogelsberger}, M., {Genel}, S., {Springel}, V., {et~al.} 2014, \mnras, 444,
  1518, \dodoi{10.1093/mnras/stu1536}

\bibitem[{{Wang} {et~al.}(2020){Wang}, {Bose}, {Frenk}, {Gao}, {Jenkins},
  {Springel}, \& {White}}]{Wang2020}
{Wang}, J., {Bose}, S., {Frenk}, C.~S., {et~al.} 2020, \nat, 585, 39,
  \dodoi{10.1038/s41586-020-2642-9}

\bibitem[{{Wechsler} {et~al.}(2002){Wechsler}, {Bullock}, {Primack},
  {Kravtsov}, \& {Dekel}}]{Wechsler2002}
{Wechsler}, R.~H., {Bullock}, J.~S., {Primack}, J.~R., {Kravtsov}, A.~V., \&
  {Dekel}, A. 2002, \apj, 568, 52, \dodoi{10.1086/338765}

\bibitem[{{Weinberger} {et~al.}(2019){Weinberger}, {Springel}, \&
  {Pakmor}}]{Weinberger2019}
{Weinberger}, R., {Springel}, V., \& {Pakmor}, R. 2019, arXiv e-prints,
  arXiv:1909.04667.
\newblock \doarXiv{1909.04667}

\bibitem[{{Zhao} {et~al.}(2003){Zhao}, {Mo}, {Jing}, \&
  {B{\"o}rner}}]{Zhao2003}
{Zhao}, D.~H., {Mo}, H.~J., {Jing}, Y.~P., \& {B{\"o}rner}, G. 2003, \mnras,
  339, 12, \dodoi{10.1046/j.1365-8711.2003.06135.x}

\bibitem[{{Z{\"u}rcher} \& {More}(2019)}]{Zurcher2019}
{Z{\"u}rcher}, D., \& {More}, S. 2019, \apj, 874, 184,
  \dodoi{10.3847/1538-4357/ab08e8}

\end{thebibliography}



\end{document}